\documentclass[a4paper]{article}
\usepackage[T1]{fontenc}
\usepackage{graphicx}
\usepackage{booktabs} 
\usepackage{subcaption}
\usepackage{caption}
\usepackage{pdflscape}
\usepackage{fancyhdr}
\usepackage{amsmath,amssymb,mathtools}
\usepackage[inline]{enumitem}

\usepackage{csquotes}
\usepackage{algorithmic}
\usepackage[table]{xcolor}
\usepackage{hyperref}
\hypersetup{colorlinks=true,
	linkcolor=magenta,
	citecolor=green,
	urlcolor=blue}
\usepackage{color}

\usepackage{listings}
\usepackage{xcolor}

\makeatletter
\renewcommand{\sectionautorefname}{\S\@gobble}
\renewcommand{\subsectionautorefname}{\S\@gobble}
\renewcommand{\subsubsectionautorefname}{\S\@gobble}
\makeatother

\usepackage[nomessages]{fp}
\newlength{\maxlen}
\newcommand{\databarred}[2][red]{%
  \settowidth{\maxlen}{7.11}%
  \addtolength{\maxlen}{\tabcolsep}%
  \FPeval\result{round(#2/7.11:4)}%
  \makebox[0pt][l]{\color{#1!30}\rule[-.05\ht\strutbox]{\result\maxlen}{.95\ht\strutbox}}%
  \makebox[\dimexpr\maxlen-\tabcolsep][r]{#2}%
}

\newcommand{\databarblue}[2][blue]{%
  \settowidth{\maxlen}{0.62}%
  \addtolength{\maxlen}{\tabcolsep}%
  \FPeval\result{round(#2/0.62:4)}%
  \makebox[0pt][l]{\color{#1!30}\rule[-.05\ht\strutbox]{\result\maxlen}{.95\ht\strutbox}}%
  \makebox[\dimexpr\maxlen-\tabcolsep][r]{#2}%
}

\newcommand{\databarpurple}[2][purple]{%
  \settowidth{\maxlen}{135}%
  \addtolength{\maxlen}{\tabcolsep}%
  \FPeval\result{round(#2/135:4)}%
  \makebox[0pt][l]{\color{#1!30}\rule[-.05\ht\strutbox]{\result\maxlen}{.95\ht\strutbox}}%
  \makebox[\dimexpr\maxlen-\tabcolsep][r]{#2}%
}

\newcommand{\databarorangeone}[2][orange]{%
  \settowidth{\maxlen}{0.000052}%
  \addtolength{\maxlen}{\tabcolsep}%
  \FPeval\result{round(#2/0.000052:4)}%
  \makebox[0pt][l]{\color{#1!30}\rule[-.05\ht\strutbox]{\result\maxlen}{.95\ht\strutbox}}%
  \makebox[\dimexpr\maxlen-\tabcolsep][r]{#2}%
}

\newcommand{\databarorangetwo}[2][orange]{%
  \settowidth{\maxlen}{0.000135}%
  \addtolength{\maxlen}{\tabcolsep}%
  \FPeval\result{round(#2/0.000135:4)}%
  \makebox[0pt][l]{\color{#1!30}\rule[-.05\ht\strutbox]{\result\maxlen}{.95\ht\strutbox}}%
  \makebox[\dimexpr\maxlen-\tabcolsep][r]{#2}%
}

\newcommand{\databarbrown}[2][brown]{%
  \settowidth{\maxlen}{35.38}%
  \addtolength{\maxlen}{\tabcolsep}%
  \FPeval\result{round(#2/35.38:4)}%
  \makebox[0pt][l]{\color{#1!60}\rule[-.05\ht\strutbox]{\result\maxlen}{.95\ht\strutbox}}%
  \makebox[\dimexpr\maxlen-\tabcolsep][r]{#2}%
}

\usepackage[sorting=none,style=numeric]{biblatex}
\usepackage{authblk}
\usepackage[margin=1in]{geometry} 

\begin{document}
\title{How the interplay between power concentration, competition, and propagation affects the resource efficiency of distributed ledgers} 

\author[1]{Paolo Barucca}
\author[2,3]{Carlo Campajola}
\author[1,3]{Jiahua Xu}

\affil[1]{Department of Computer Science, University College London, 66-72 Gower Street, WC1E 6EA London, United Kingdom}
\affil[2]{Institute of Finance and Technology, University College London, Gower Street, WC1E 6BT London, United Kingdom}
\affil[3]{DLT Science Foundation, London, United Kingdom}



%
%
%

\maketitle              
\begin{abstract}
Forks in the Bitcoin network result from the natural competition in the blockchain's Proof-of-Work consensus protocol. Their frequency is a critical indicator for the efficiency of a distributed ledger as they can contribute to resource waste and network insecurity.
We introduce a model for the estimation of natural fork rates in a network of heterogeneous miners as a function of their number, the distribution of hash rates and the block propagation time over the peer-to-peer infrastructure. 
Despite relatively simplistic assumptions, such as zero propagation delay within mining pools, the model predicts fork rates which are comparable with the empirical stale blocks rate. 
In the past decade, we observe a reduction in the number of mining pools approximately by a factor 3, and quantify its consequences for the fork rate, whilst showing the emergence of a truncated power-law distribution in hash rates, justified by a rich-get-richer effect constrained by global energy supply limits. 
We demonstrate, both empirically and with the aid of our quantitative model, that the ratio between the block propagation time and the mining time is a sufficiently accurate estimator of the fork rate, but also quantify its dependence on the heterogeneity of miner activities. 
We provide empirical and theoretical evidence that both hash rate concentration and lower block propagation time reduce fork rates in distributed ledgers. 
Our work introduces a robust mathematical setting for investigating power concentration and competition on a distributed network, for interpreting discrepancies in fork rates---for example caused by selfish mining practices and asymmetric propagation times---thus providing an effective tool for designing future and alternative scenarios for existing and new blockchain distributed mining systems.

\end{abstract}

\section{Introduction}
The Bitcoin protocol \cite{Nakamoto2008Bitcoin:System}, introduced in 2008, is the first and most famous blockchain-based system operating at scale. It consists of an append-only, public and permission-less ledger---the blockchain, which records transactions of the Bitcoin cryptocurrency between pseudonymous addresses. While the ledger is public and any person with an internet connection and the publicly available node software can read its contents, write access to the ledger is secured by a cryptographic puzzle whose solution involves the local inversion of the SHA-256 hash function \cite{Lilly2004DeviceHashing}. 

This puzzle can only be solved through brute-force calculation, and the winner of this race gains the right to append a new set of valid transactions (a block) to the most recent block and to create a predetermined amount of new Bitcoins that they can take as a reward. Engaging in this process is known as Proof-of-Work (PoW) mining, and its operators are called miners. The puzzle's difficulty is periodically and automatically adjusted by the protocol so that on average a solution is found every 10 minutes, which is the target block time of the Bitcoin ledger. 
Given this mechanism, write rights are also permission-less in principle: there is no central authority that has the right to censor or deny write access, and overwriting is also not possible thanks to the protocol's reliance on Merkle trees and chained hashes.

However, due to risk management and the strong economies of scale in technology markets \cite{Arthur1994IncreasingEconomy}, the competition for write access favors the players who have enough economic resources to acquire large numbers of sophisticated, purpose-built machines---known as Application-Specific Integrated Circuits or ASICs---which severely outperform commercial CPUs and GPUs in the solution of the PoW puzzle. This economic mechanism naturally pushes retail miners out of the mining market, as they end up having a negligible probability of successfully mining a block with respect to the most equipped and capitalized entities. Large mining operators also aggregate into even larger mining pools as a way to smooth out their reward payout and share the risk associated with the randomness of the PoW protocol.

This emerging concentration leads to a strongly heterogeneous distribution of hash rates, i.e. the hashes per second (H/s) that miners are able to generate. At the time of writing, 2 mining pools produce over 50\% of new Bitcoin blocks\footnote{Source: \url{https://www.blockchain.com/explorer/charts/pools}, accessed 09 February 2024}, which is a proxy for their share of the total hash rate that is being committed to Bitcoin mining. 
This is concerning for several reasons: first, if a single entity could control the absolute majority of mining power, it would be able to run a 51\% attack \cite{Avarikioti2019BitcoinMajority}, i.e. to write false information on the blockchain while always producing the longest chain, which is the standard rule by which nodes determine the \enquote{validated} blockchain; second, concentration distorts the fees market, which Bitcoin users have to pay to incentivize the miners to include their transaction in the limited space of the next block; third, it puts major miners in a censorship position, as they could arbitrarily delay the inclusion of specific transactions on the blockchain.

In this paper, we explore the impact of the heterogeneity of the hash rate distribution on the probability of generating soft forks. These situations arise when two miners solve the PoW puzzle almost simultaneously, and then have to compete to be the first to broadcast their block to the majority of other network nodes. This competition is due to the network latency, which in Bitcoin leads to an average propagation time of about 2 seconds to reach 90\% of the network (see \autoref{sec:data}). 
A soft fork \cite{Decker2013InformationNetwork,eyal2014majority}
can be resolved by the competition of the two versions of the block. The block which eventually is not included in the longest chain becomes an \textit{orphaned} block, and its content will not be considered validated.
Soft forks are an inefficiency of the Bitcoin protocol that leads to wasted computational resources (and thus energy), increasing the cost of network operations and its environmental impact to maintain a given level of security. 

\begin{figure}[t]
\centering
\includegraphics[width=\linewidth]{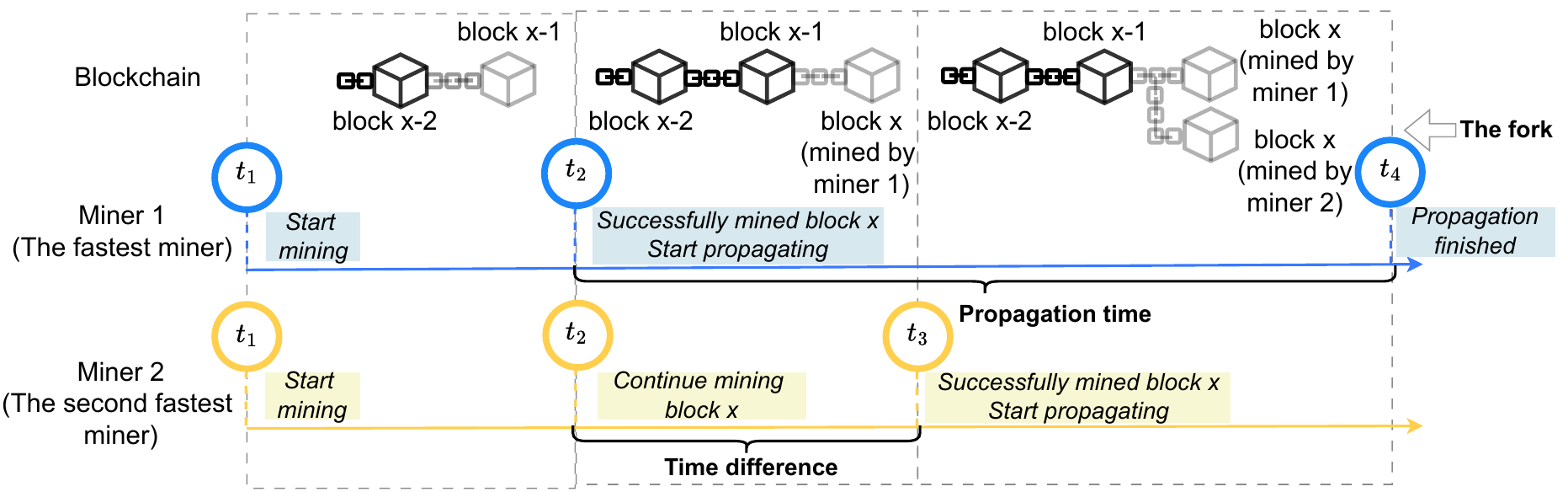}
\caption{\footnotesize 
To simplify the dynamics, we assume that all miners start mining at $t_1$. At $t_2$ Miner 1 solves the PoW puzzle for block $x$ and starts broadcasting it to the network. By $t_4$, the message has spread to most of the nodes. However, Miner 2 successfully mines the block at $t_3$, which is later than $t_2$ and earlier than $t_4$. Therefore, based on the definition above, the propagation time is $t_4-t_2$, and the fork is present between time  $t_4$ and $t_3$.}
\label{fig:ForkGeneration}
\end{figure}



We base our modeling approach on previous work by Fadda et al.\cite{Fadda2022ConsensusNetworks}, who investigate the impact of network structure on the generation of soft forks. Our intent is to relax their assumption that all miners are equal, i.e. have the same hash rate, accounting for a more realistic hash rate distribution, while keeping our solution fully analytical to avoid cumbersome simulations. 
We take literature estimates for the frequency of soft forks in the Bitcoin blockchain \cite{Decker2013InformationNetwork}, and compare it with the results of our modeling approach, which takes as input the number of miners and the parameters of the hash rate distribution. 

Decker et al. \cite{Decker2013InformationNetwork} first investigated the fork generation from a network propagation perspective verifying that the primary cause of forks is indeed block propagation delay. The authors quantified the probability of the block mining time and fork generation with the assumption that the computational power is uniformly distributed. Remarkably, according to their simple model, the probability of forks was 1.78\%, while their empirical data suggested a probability of 1.69\%. 
Neudecker et al. \cite{Neudecker2019} monitored empirical data and observed that the probability of a block being incorporated into the main chain has a roughly linear relationship with the time that block has been disseminated prior to the competing block.  Shahsavari et al. \cite{Shahsavari2019PerformanceProtocol} extended the fork probability equation further by introducing new factors, such as block size and network conditions, which determine the propagation time. 
They assumed the block arrivals as a homogeneous Poisson process and found that smaller block sizes and higher bandwidth have the effect of decreasing the fork rate, finding good agreement with empirical fork rates.

Tessone et al. \cite{Tessone2021StochasticConsensus} proposed a stochastic model and performed numerical simulations to investigate the dynamics of blockchain-based consensus. They found that the concentration of mining power, measured by its Gini index, tends to decrease as network latency increases. Numerical simulations conducted by Liu et al. \cite{Liu2021} showed that the fork rate increases with the number of miners, $N$.

As for the theoretical approach which is at the basis of our present contribution, Fadda et al. \cite{Fadda2022ConsensusNetworks} estimated the probability of forks by computing analytically the probability that the time difference of the two fastest miners is lower than the propagation delay needed to reach the whole network. Based on the assumption of homogeneous miners, they find that their model tends to overestimate the frequency of soft forks. 

Finally, studies on forks are closely related to the growing literature on strategic mining behavior, like selfish mining \cite{Eyal2014MajorityVulnerable}, stubborn mining \cite{Nayak2016StubbornAttack} or other practices \cite{Liu2018OnN-attackers}. Eyal and Sirer \cite{Eyal2014MajorityVulnerable} originally pointed out that the Bitcoin mining protocol is vulnerable to a potential attack where miners withhold their mined blocks, exploiting network delay and hash rate concentration to gain unfair advantages in mining sequences of blocks, rather than single blocks. While detection of the practice has been scarce, recent research \cite{Li2022TwistedMining} has devised a statistical test that finds anomalous activity by miners that would be compatible with strategic mining.



In the following, we show the emergence of mining pools in the past decade and are able to interpret their macroscopic effects on the evolution of propagation time and fork rates. We demonstrate that while the ratio between the block propagation time and the mining time is a sufficiently accurate estimator of the fork rate, yet non-linear effects arise due to the heterogeneity of miner activity that require more sophisticated estimators.
Further, we show the emergence of a truncated power-law distribution in hash rates and its consequences for the fork rate.
Overall, we provide a quantitative model for investigating and interpreting the consensus dynamics on a distributed network and for designing future and alternative scenarios for existing and new blockchain mining systems.

\begin{figure}[!ht]
\centering
\includegraphics[width=\linewidth]{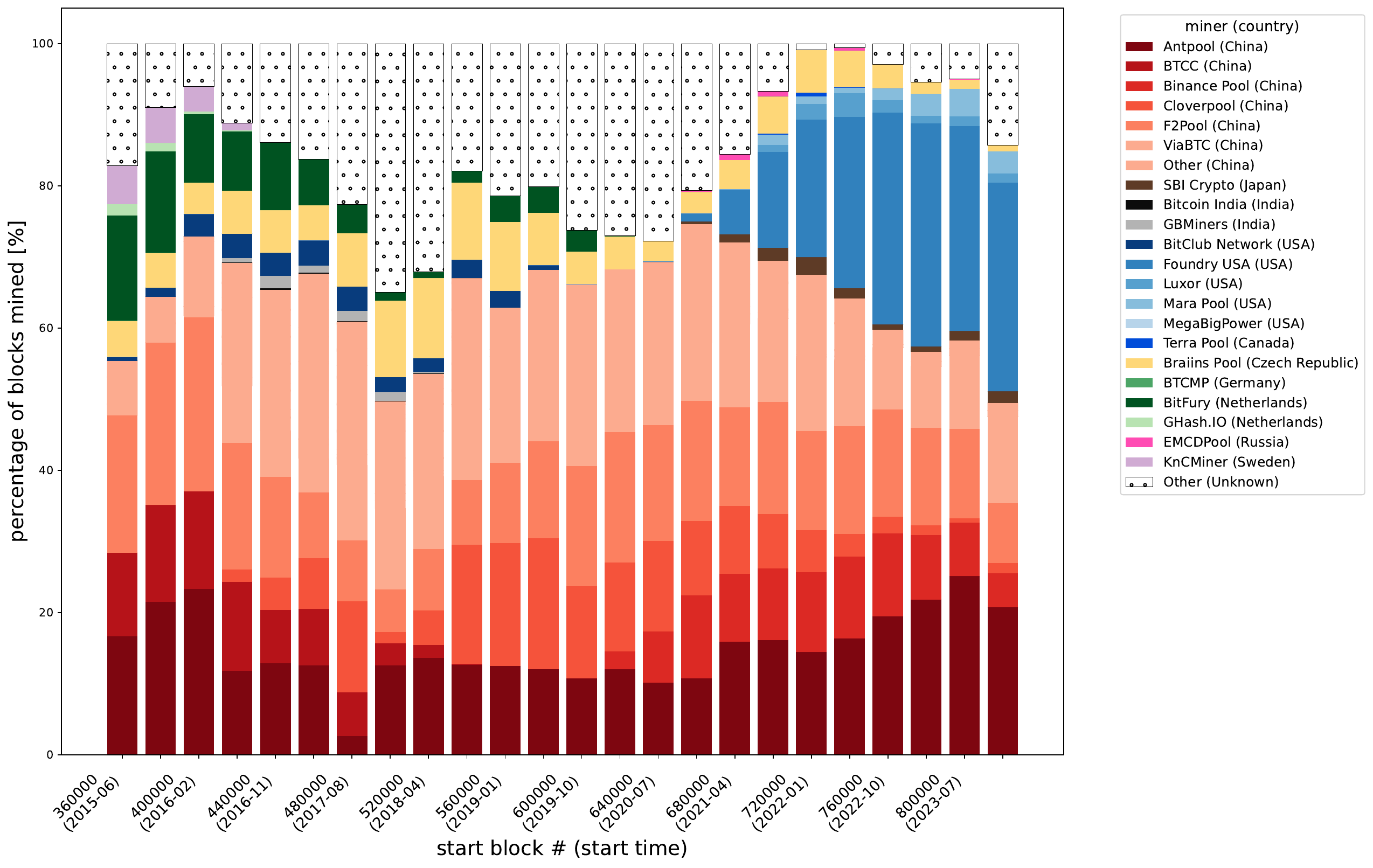}
\caption{
Share of blocks mined among miners in selected observation periods. Unknown miners - for which no specific signature is available - are in dotted white, and the other minority China miners are in dotted light red.}
\label{fig:miner_stacked}
\end{figure}

\section{Results}

From \autoref{fig:miner_stacked}, we detect a clear trend of mining power concentration throughout the last decade. The total number of active miners was around 100 before 2019, with the hash rate share of the largest miner being as low as 13\% at one time (mid 2017). At the time of writing, the miner number decreased to around 35, with the largest one accounting for a 35\% share.
\autoref{fig:miner_stacked} also shows the dominance of mining pools based in China, such as Antpool and BTCC. Since 2022, some of those pools phased out---likely due to the China's Bitcoin Mining Ban in 2021---and US pools such as Foundry USA started to take lead\footnote{\url{https://www.bbc.co.uk/news/technology-58896545}}. We report the complete summary statistics in the Supplementary Material \autoref{subtab:hash_empirical}.

\label{sec:simulation}

\begin{figure}[!h]
    \begin{subfigure}[t]{0.328\textwidth}
    \includegraphics[height=0.17\textheight, trim=0 8 0 0, clip]{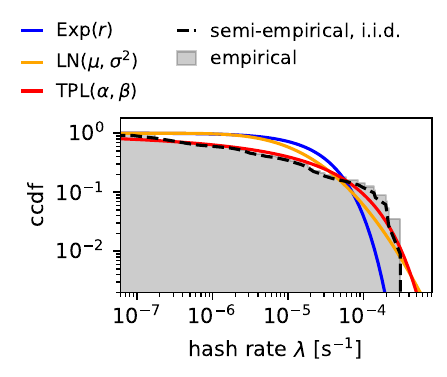}
    \vskip -0.2cm
    \caption{block 420000-440000}
    \end{subfigure}
    \hfill
    \begin{subfigure}[t]{0.328\textwidth}
    \includegraphics[height=0.17\textheight, trim=0 8 0 0, clip]{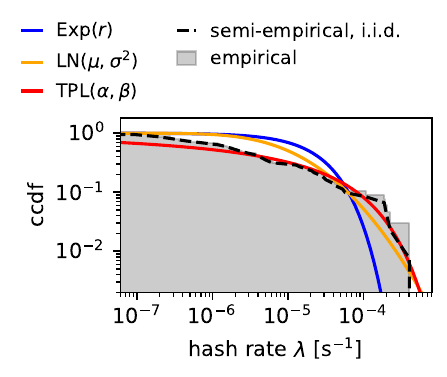}
    \vskip -0.2cm
    \caption{block 500000-520000}
    \end{subfigure}
    \hfill
    \begin{subfigure}[t]{0.328\textwidth}
    \includegraphics[height=0.17\textheight, trim=0 8 0 0, clip]{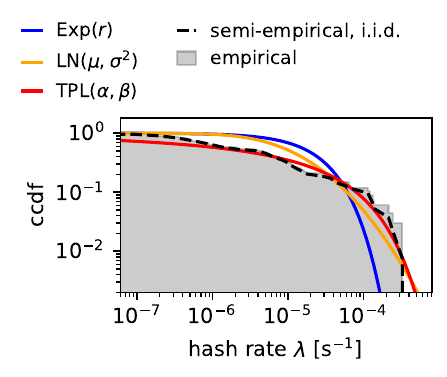}
    \vskip -0.2cm
    \caption{block 580000-600000}
    \end{subfigure}


    \begin{subfigure}[t]{0.328\textwidth}
    \includegraphics[height=0.12\textheight, trim=0 8 0 50, clip]{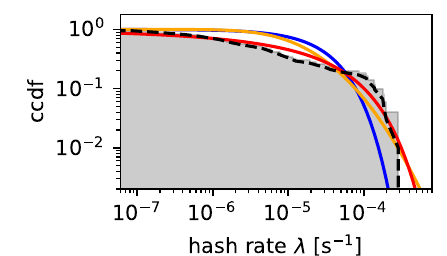}
    \vskip -0.2cm
    \caption{block 660000-680000}
    \end{subfigure}
    \hfill
    \begin{subfigure}[t]{0.328\textwidth}
    \includegraphics[height=0.12\textheight, trim=0 8 0 50, clip]{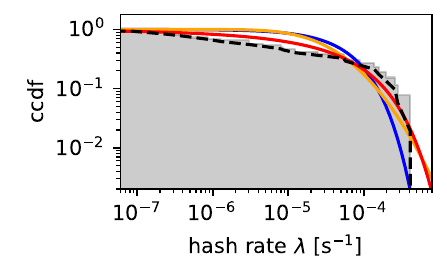}
    \vskip -0.2cm
    \caption{block 740000-760000}
    \end{subfigure}
    \hfill
    \begin{subfigure}[t]{0.328\textwidth}
    \includegraphics[height=0.12\textheight, trim=0 8 0 50, clip]{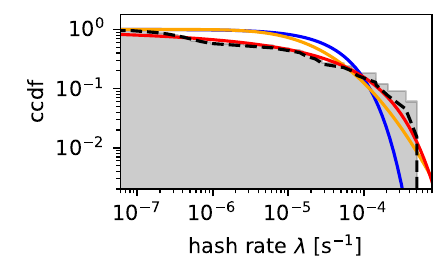}
    \vskip -0.2cm
    \caption{block 820000-840000}
    \label{fig:lastperiod}
    \end{subfigure}

    \caption{Complementary cumulative distribution function (ccdf) of hash rates' empirical distribution and null distributions---exponential, log normal, and truncated power law---fitted through the method of moments \eqref{eq:mom} in selected observation periods.}
\label{fig:ccdf_emp}
\end{figure}

This strong heterogeneity is confirmed when looking at the cross-sectional distribution of hash rates $\lambda_i$. We show this distribution at several points in time in \autoref{fig:ccdf_emp}, and observe that as time passes it gets more and more skewed towards the right tail, indicating the emergence of \enquote{mega miners} dominating the mining of the Bitcoin network. We compare the empirical probability to a set of fitted null distributions, namely the exponential $\lambda_i \sim \mathrm{Exp}(r)$, the log-normal $\lambda_i \sim \mathrm{LN}(\mu, \sigma^2)$ or the truncated power law $\lambda_i \sim \mathrm{TPL}(\alpha, \beta)$, as well as a semi-empirical Bayesian posterior that accounts for the potential difference between the estimator we use (i.e. the share of mined blocks) and the actual share of hashing power. 
The displayed distributions all consider i.i.d. hash rates; the null distributions are fitted using the method of moments \autoref{eq:mom}, while the semi-empirical one was computed according to \autoref{eq:plamb_iid}. We further motivate our choice of null distributions and the estimation techniques for their parameters in the Methods section.

Among the null distributions, the log-normal and truncated power law are the closest fit, both able to capture the right tail---with the truncated power law distribution slightly more so---better than exponential. However, the semi-empirical distribution is better at capturing the tail thickness and cut-off. These facts are well-known from a growing body of literature on the emerging inequality in cryptocurrencies, however the impact of heterogeneity in mining power on the security and efficiency of the network has not been studied with a mathematical model so far. The main goal of our work is to develop such model, and quantify the impact of hashrate and geographical concentration on the emergence of soft forks. Having such a model not only allows to understand the consequences of centralization but also to calculate the implied primitives of the model based on the available data on forks and propagation times, which can then be used for anomaly detection. In the following paragraphs we introduce a minimalistic model and describe how it can be used to connect fork rates, network latency and inequality in hashrates.

\paragraph{Modeling consensus propagation in a heterogeneous network.}
\label{sec:model}
We assume that the number of successful mining events occurring in a fixed time period is Poisson distributed, as the continuous time limit of a Bernoulli process with low enough success probability. Therefore, in our model, the time to successfully mine a block is exponentially distributed, and the probability of a block being mined at time $t$ (assuming $t_1 = 0$) for miner $i$ is
\begin{equation}
p_i(t)=\frac{\lambda_i}{e^{\lambda_i t}},
\label{eq:p_i(t)}
\end{equation}
where $\lambda_i$ represents the hash rate of miner $i$, i.e. the expected number of blocks per unit of time the miner can independently mine (i.e. without the influence of competition).
The minimum time for any miner to succeed mining a block, $t_{\min}$, is the first order statistic $t_{(1)}$ of the system of miners, which has probability distribution function
\begin{equation}
p_{(1)}(t \vert \{\lambda_i\})=\sum_{i=1}^N \left( p_i(t) \prod_{j \neq i}^N S_j(t) \right) ,
\label{eq:p(1)_1}
\end{equation}
where  $\{\lambda_i\} = \{\lambda_1, \dots, \lambda_N\}$, $N$ is the number of miners, and $S_j(t) = \int_t^\infty p_j(s) ds = e^{-\lambda_j t}$ is the probability that miner $j$'s mining time exceeds $t$. 
Plugging exponential distributions in the equation, it reads
\begin{equation}
p_{(1)}(t \vert \{\lambda_i\})= \Lambda e^{-\Lambda \cdot t }
\label{eq:p(1)_2} ,
\end{equation}
where $\Lambda = \sum_i \lambda_i$ represents the aggregate mining power of the network. 
It is then clear that the expected minimum mining time in this setup reads
\begin{equation}
\left\langle t_{\text {min }}\right\rangle =\tfrac{1}{\Lambda}
\label{eq:block_time}
\end{equation}

that, for the Bitcoin network, is kept at approximately 10 minutes by the adjustment of the PoW puzzle difficulty parameter.
To estimate the probability of a soft fork to occur, we need to obtain the distribution for the time difference between the two fastest miners, which can be obtained from the joint distribution of the first and second order statistics:
\begin{equation}
\begin{split}
p\left(t, t' \vert \{\lambda_i\} \right)&= \sum_{i \neq j} \theta\left(t'-t\right) p_i(t) p_j\left(t'\right) \prod_{k \neq i, j}^N S_k(t')
\nonumber\\
&=
\sum_{i \neq j} \frac{\theta\left(t'-t\right) \lambda_i \lambda_j}{e^{\lambda_i\left(t-t'\right)+\sum_k \lambda_k t'}} 
\end{split}
\label{eq:p_t,t'}
\end{equation}
where $t'$ is the second shortest mining time and $\theta$ is the Heaviside step function, equal to 1 for $t'>t$ and 0 otherwise.
%
The distribution of time differences $\Delta = t'-t$ can then be derived as
\begin{align}
p(\Delta=t'-t \vert \{\lambda_i\}) =\int d t d t' \delta ( \Delta-t'+t ) p (t, t' \vert \{\lambda_i\} )
=\frac{\sum_i^N \frac{\lambda_i \sum_{j \neq i}\lambda_j}{e^{\Delta \sum_{j \neq i} \lambda_j}}}{\sum_i^N\lambda_i}  
,
\label{eq:cond_pdelta}
\end{align} 
where $\delta(x)$ is the Dirac delta function, that selects only the values of $t'$ and $t$ such that their difference equals $\Delta$. 

The probability of a time gap between the two fastest miners smaller than or equal to $\Delta_0$ can be computed as $\Delta$'s cumulative probability,
\begin{align}
    C(\Delta_0 \vert \{\lambda_i\}) = \int_0^{\Delta_0} p(\Delta \vert \{\lambda_i\}) dt
    = \frac{\sum_i^N \int_0^{\Delta_0}
    \frac{\lambda_i \sum_{j \neq i}\lambda_j}{e^{\Delta \sum_{j \neq i} \lambda_j}} d\Delta}{\sum_i^N\lambda_i}
    =
    \frac{\sum_i^N 
    \lambda_i \left( 1- \frac{ 1}{e^{\Delta_0 \sum_{j \neq i} \lambda_j}} \right)
    }{\sum_i^N\lambda_i}
    = 1-  \tfrac{\sum_i^N \tfrac{ \lambda_i}{e^{\Delta_0 \sum_{j \neq i} \lambda_j}}}{\sum_i^N\lambda_i}.
\label{eq:cdelta_cond}
\end{align}
Since $C(\Delta_0 \vert \{\lambda_i\})$ is the probability that a soft fork occurs, $p(\Delta_0 \vert \{\lambda_i\})$ represents the sensitivity of fork rate to propagation time $\frac{d C(\Delta_0 \vert \{\lambda_i\})}{d \Delta_0}$. Specifically at $\Delta = 0$, this sensitivity can be simplified as
\begin{equation}
p(0 \vert \{\lambda_i\}) = 
\frac{\sum_i^N  \lambda_i \sum_{j \neq i}\lambda_j}{\sum_j^N\lambda_j}
=
\frac{
\sum_i^N 
\left[
\lambda_i (\sum_j^N\lambda_j  - \lambda_i)
\right]}{\sum_j^N\lambda_j}
= \left(\sum_i^N 
\lambda_i \right) \left[
1 - \sum_i^N\left(\frac{\lambda_i}{\sum_j^N\lambda_j} \right)^2
\right]
\end{equation}
where $\sum_i^N\left(\frac{\lambda_i}{\sum_j^N\lambda_j} \right)^2 \eqqcolon {\it HHI} \in (0, 1]$ is the Herfindahl-Hirschman Index (HHI) of the mining network, describing the level of hash rate concentration: higher HHI represents a more concentrated distribution of hash rates among miners, whereas a lower value signifies a more equal distribution. 
%
At sufficiently small $\Delta_0$, $C(\Delta_0 \vert \{\lambda_i\})$ can be estimated with its first-order Taylor series approximation,
\begin{equation}
    C(\Delta_0 \vert \{\lambda_i\}) \approx \Delta_0 \cdot p(0 \vert \{\lambda_i\}) = \Delta_0 \cdot \Lambda \cdot (1- {\it HHI}) 
    \label{eq:taylorc}
\end{equation}

Eq. \eqref{eq:cdelta_cond} represents conditional cumulative probability: we can then take an ergodic approach and consider the average over different realizations of hash rates, to obtain the unconditional one
\begin{align}
\label{eq:cdelta}
    C(\Delta_0)
    = 1- \int_D p(\{\lambda_i\})  
    \left( \tfrac{\sum_i^N \tfrac{ \lambda_i}{e^{\Delta_0 \sum_{j \neq i} \lambda_j}}}{\sum_i^N\lambda_i}\right) d\{\lambda_i\}.
\end{align}%
where $D \subset \mathbb{R}_+^N$ is the $N$-dimensional hyperplane of $\{\lambda_i\}$'s viable range.

%

%
When $\forall i\neq j, \lambda_i \perp \lambda_j$, we can write out $C(\Delta_0)$ as
\begin{align}
\label{eq:cdelta_id}
C(\Delta_0) = 1-  \int_0^\infty  
    \left[
        \sum_i^N 
    \left(
     \int_0^{\infty}\tfrac{\lambda_i p(\lambda_i)d{\lambda_i}}{e^{x\lambda_i}} 
    \prod_{j \neq i}
    \int_0^{\infty} \tfrac{p(\lambda_j)d{\lambda_j}}{e^{(\Delta+x)\lambda_j}}
    \right)
    \right] dx .
\end{align}
The derivation of Eq. \eqref{eq:cdelta_id} can be found in Supplementary Material.
When we impose the additional assumption of identical distribution, i.e. $\{\lambda_i\}$ are i.i.d., we get
\begin{align}
\label{eq:cdelta_iid}
C(\Delta_0) = 1-  N \int_0^\infty  
    \left[\left(
     \int_0^{\infty}\tfrac{\lambda p(\lambda)d{\lambda}}{e^{x\lambda}}\right) 
    \left(
    \int_0^{\infty} \tfrac{p(\lambda)d{\lambda}}{e^{(\Delta_0+x)\lambda}}
    \right)^{N-1}
    \right] dx .
\end{align}

\begin{figure}[tb]
\centering
    \begin{subfigure}[t]{0.48\textwidth}
    \includegraphics[height=0.65\linewidth,trim=0 0 0 0, clip]{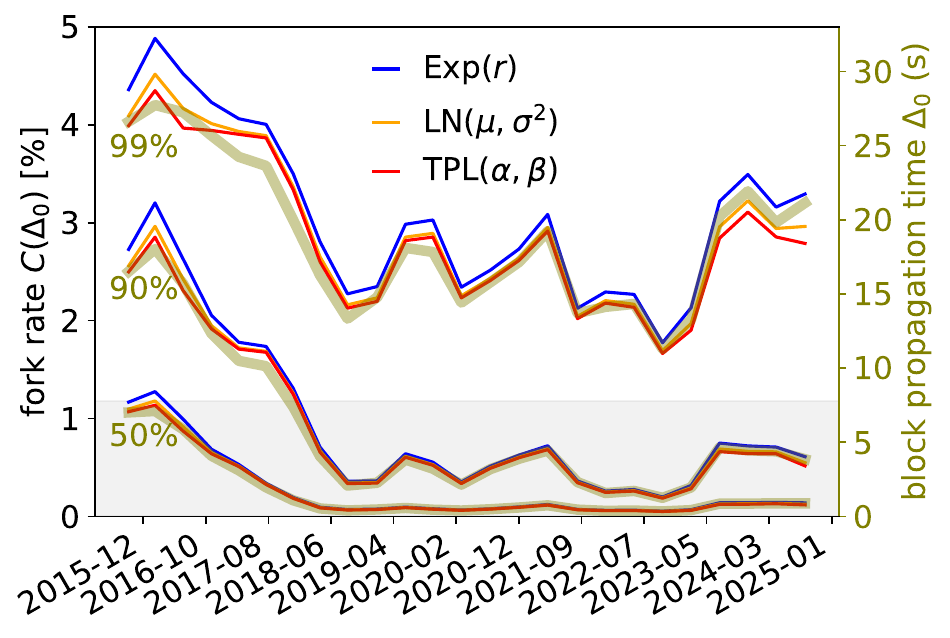}
    \caption{Model-estimated fork rates under different null distribution assumptions and at different levels of block propagation time $\Delta_0 \in \{\Delta_0^{(50)}, \Delta_0^{(90)}, \Delta_0^{(99)} \}$} 
    \label{fig:forkrate_full}
    \end{subfigure}
    \hfill
    \begin{subfigure}[t]{0.48\textwidth}
    \includegraphics[height=0.65\linewidth,trim=0 0 0 0, clip]{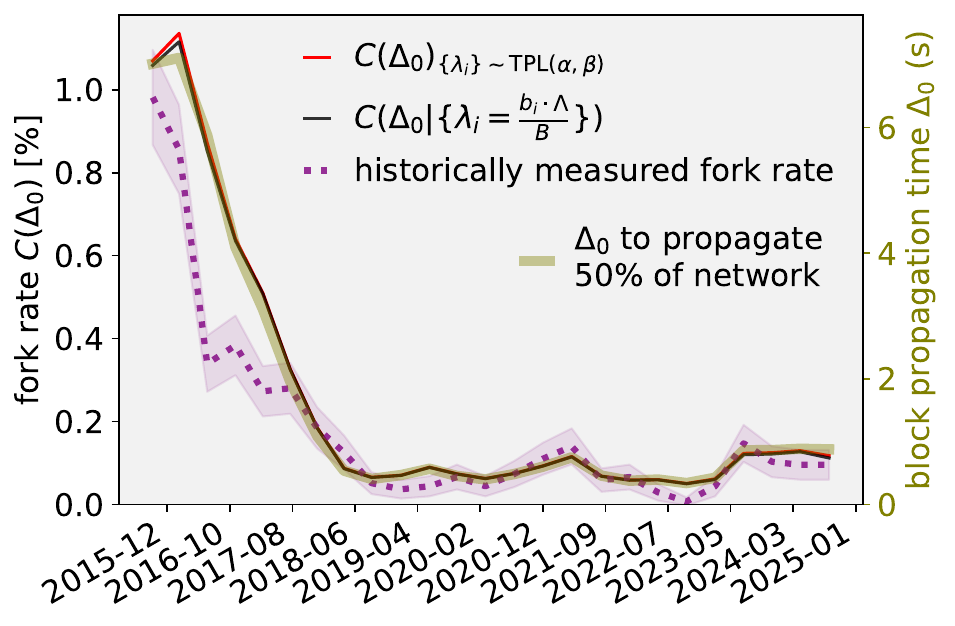}
    \caption{Model-estimated fork rates using $\text{TPL}(\alpha, \beta)$ versus using empirical $\{\lambda_i\}$, for $\Delta_0 = \Delta_0^{(50)}$ network propagation time, compared with historically measured fork rates} 
    \label{fig:forkrate_small}
    \end{subfigure}
\caption{Time series of historically measured fork rates (dotted line) compared with model-estimated fork rates under the assumption of various distributions when different values of propagation time are used. The shaded area around the historical rates represents 90\% confidence band.
}  
\label{fig:forkrate_time_series}
\end{figure}


%

%

\subsection*{Implied versus empirical fork rates}

The main advantage of having an analytical formulation of fork generation from mining and network data is that we can use it to validate its inputs based on the empirical frequency with which forks occur. In other words, we can identify under which combinations of $\Delta_0$ and $p(\{ \lambda_i \})$ the fork rate is compatible with the empirically measured one, and compare them with the block propagation times and the share of mined blocks respectively. In \autoref{fig:forkrate_time_series}, we compare model-estimated fork rates using fitted parameters (\autoref{subtab:hash_dis}) with historical fork rates (see \autoref{subtab:hash_empirical}). To choose the threshold time $\Delta_0$, we use the estimates of the time it takes for a block to reach 50\%, 90\% or 99\% of the Bitcoin network (\autoref{fig:forkrate_full}), which we name $\Delta_0^{(50)}$, $\Delta_0^{(90)}$ and $\Delta_0^{(99)}$ respectively. We observe that exponentially distributed $\lambda_i$s yield the highest fork rate, followed by log-normally and then TPL-distributed hashrates, although the gap between the latter two is rather negligible. 

\begin{figure}[!t]
\centering

    \includegraphics[height=0.2\textheight, trim=0 8 0 0, clip]{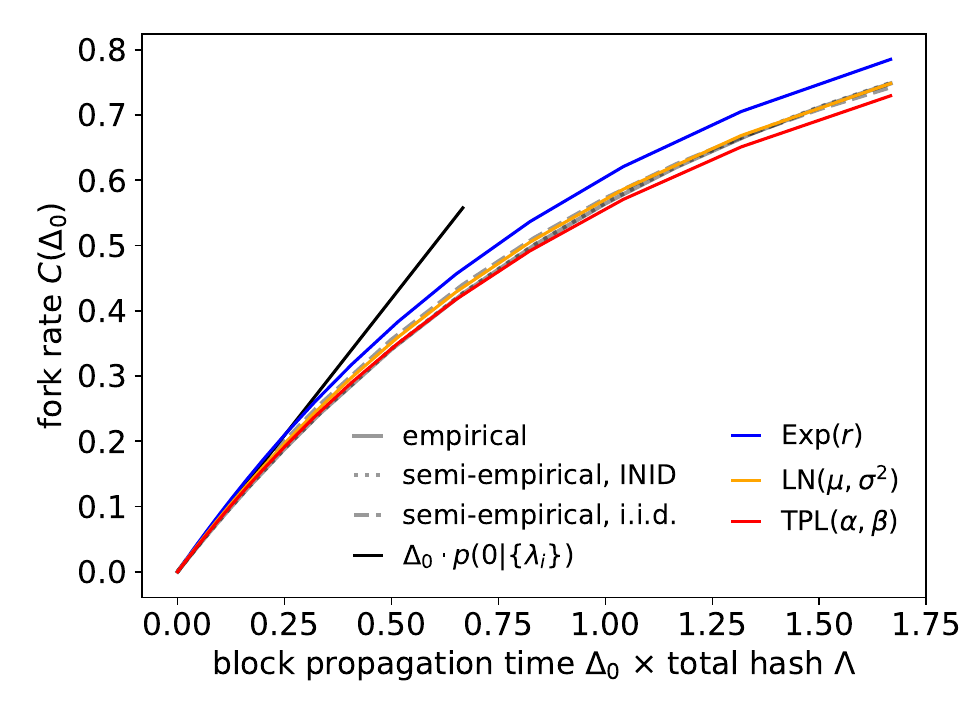}
    
    
\caption{Fork rate estimation at different block propagation times, comparing the one obtained numerically via the empirical ccdf, the semi-empirical Bayesian estimates and the null distributions.}
\label{fig:id_vs_iid}
\end{figure}

We show in \autoref{fig:forkrate_small} that the model-predicted fork rate is largely consistent with the empirically estimated fork rate when $\Delta_0 = \Delta_0^{(50)}$. We also find that there is close to no difference between the estimated fork rate using a fitted distribution such as $\text{TPL}(\alpha,\beta)$ versus numerically solving the model with the empirically measured $\{\lambda_i = \frac{b_i \cdot \Lambda}{b}\}$, suggesting a high goodness of fit.



\autoref{fig:forkrate_small} would suggest that the choice of null distribution has a marginal impact on the determination of the fork rate. To better investigate this, in \autoref{fig:id_vs_iid} we compare the estimated fork rate under several distributional assumptions, applying Eq. \eqref{eq:cdelta_id}
and \eqref{eq:cdelta_iid}, and varying the characteristic time $\tau = \Delta_0 \Lambda = \frac{\Delta_0}{\langle t_{\min} \rangle}$, i.e. the ratio between the block propagation time and the expected time to mine a new block.
We can see that the difference between fork rates is barely distinguishable for $\tau < 0.4$, or $\Delta_0 < 240$ [s] in the case of Bitcoin, which is generally much larger than the real block propagation times, and that the linear approximation of \autoref{eq:taylorc} is largely valid over realistic ranges of $\tau$. It is however good to notice that second-order effects start to become relevant at higher values of $\tau$: these values are obviously rather unrealistic for Bitcoin, but may be attainable in faster PoW blockchains like Litecoin or Dogecoin, that operate with average block times of 2.5 and 1 minute respectively. At large $\Delta_0$, we see that the estimated fork rate is slightly higher when assuming $\lambda_i$s are independent but not identically distributed than under the i.i.d. assumption. 

%

\begin{figure}[tb]
\centering
    \begin{subfigure}[t]{0.48\textwidth}
    \includegraphics[height=0.65\linewidth,trim=0 0 0 0, clip]{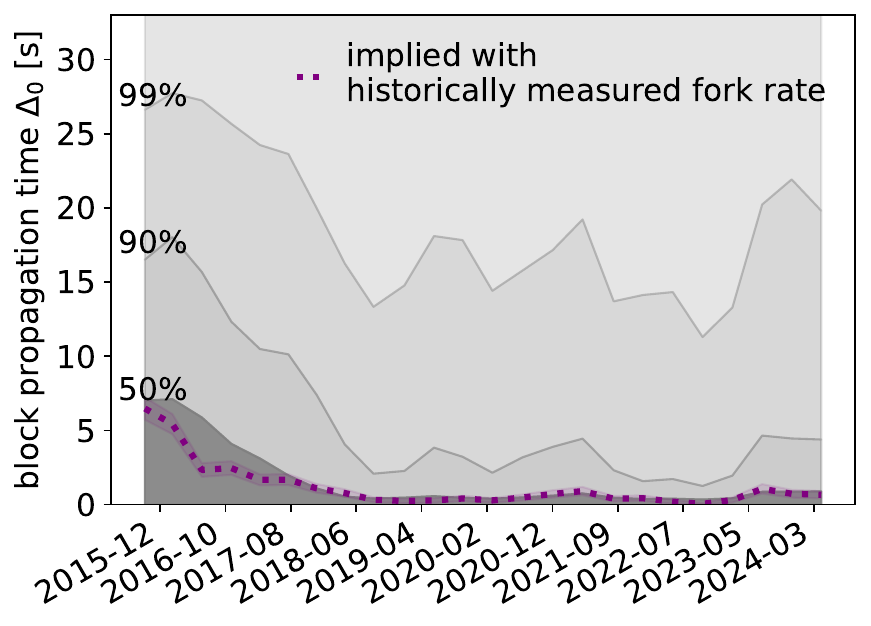}
    \caption{implied $\Delta_0$ as $\tfrac{\text{historically measured fork rate}}{\Lambda \cdot (1- {\it HHI})}$ compared with empirical block propagation time: to reach 50\%, 90\%, 99\% of the network, respectively}
    \label{fig:forkrate_ratio}
    \end{subfigure}
    \hfill
    \begin{subfigure}[t]{0.48\textwidth}
    \includegraphics[height=0.65\linewidth,trim=0 0 0 0, clip]{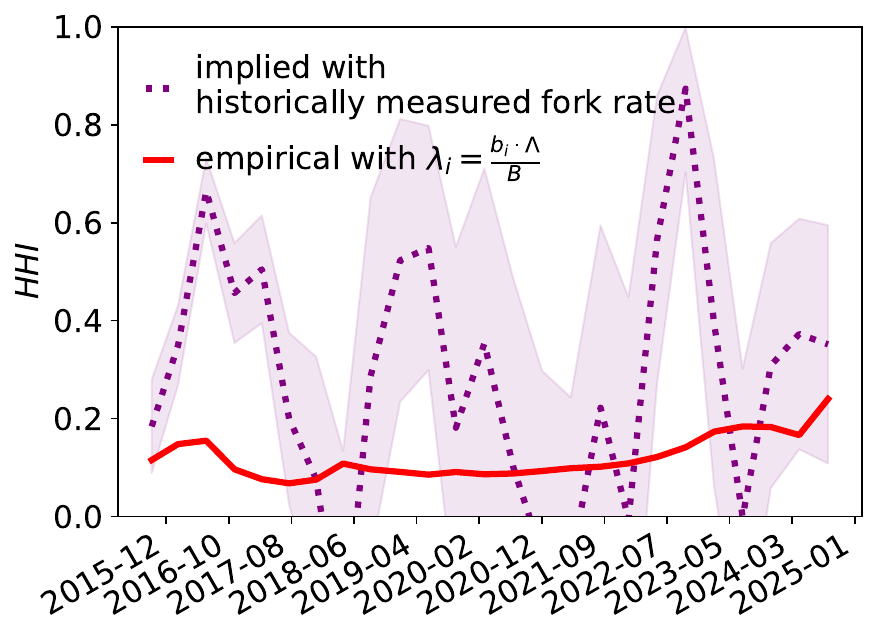}
    \caption{implied $\it HHI$ as $1 - \tfrac{\text{historically measured fork rate}}{\Delta_0 \cdot \Lambda}$ compared with empirical $\it HHI$ using empirical $\lambda_i = \frac{b_i \cdot \Lambda}{B}$} 
    \label{fig:forkrate_hhi}
    \end{subfigure}
\caption{Time series of implied $\Delta_0$ and implied $\it HHI$ versus their empirical values.
}  
\label{fig:ratio_hhi}
\end{figure}

These results suggest that the linear approximation of \autoref{eq:taylorc} is generally good enough to characterize the relation between $\Delta_0$, $p(\{\lambda_i\}) $ and the fork rate. 
A very interesting result can be obtained by inverting this equation to calculate the \enquote{\textit{implied} $\Delta_0$} and the \enquote{{implied HHI}}, i.e. the values of $\Delta_0$ and \textit{HHI} that would produce the empirical fork rates. In \autoref{fig:forkrate_ratio} we show the implied $\Delta_0$ given the historical fork rate and the measured HHI from the empirical $\{\lambda_i\}$, compared with the historical block propagation times $\Delta_0^{(50)}$, $\Delta_0^{(90)}$, and $\Delta_0^{(99)}$ of the network. We see how the implied $\Delta_0$ is typically smaller than or equal to the median time it takes to broadcast a new block to Bitcoin nodes: this would suggest that miners have a connectivity within the network that is better than most nodes, i.e. they are among the first 50\% of nodes that hear about a new mined block, thus avoiding the generation of forks.

We also compute the ${\it HHI}$ implied by the historically measured fork rate (\autoref{fig:forkrate_hhi}) assuming that $\Delta_0 = \Delta_0^{(50)}$, and compare its time series with the ${\it HHI}$ computed using empirically observed block mining data (Eq. \eqref{eq:emp_hhi}). We see that the implied ${\it HHI}$ is significantly higher than the empirically measured ${\it HHI}$ in most of the sample; however, we also see how it sometimes drops to much lower and even negative values, which would challenge the interpretation of the indicator as an \textit{HHI}. This is a consequence of fixing $\Delta_0 = \Delta_0^{(50)}$: a negative implied \textit{HHI} means that the $\Delta_0$ being used is too large.

Given the above results, it now becomes clear that the overestimation of the model-implied fork rates seen in \autoref{fig:forkrate_small} can likely be explained by (a combination of) two factors:
\begin{enumerate*}[label={(\roman*)}]
    \item a lower actual block propagation time ($\Delta_0$), and/or
    \item a higher power concentration, or lower competition, (${\it HHI}$)
\end{enumerate*}
than the value that can be estimated by tracking block signing by miners. 
For the years before 2019 in particular, it may be because the most powerful miners around that period were geographically concentrated in one area, e.g. China (see \autoref{fig:miner_stacked}), and the communication between them was faster than between any two average nodes, hence lower actual $\Delta_0$ than the one we used to calculate fork rate. This would suggest that the Bitcoin peer-to-peer infrastructure might present core-periphery features, with a strongly connected core of nodes including the majority of miners, and a loosely connected periphery of nodes that do not contribute to the mining process. This centrality of miners has already been reported in economic analyses of transaction networks \cite{Campajola2022ThePlatforms, Makarov2021BlockchainMarket}, and these results would be consistent with a similar result for the peer-to-peer overlay. An alternative explanation would be that there has been sporadic collusion between miners, which would result in lower actual ${\it HHI}$.



\begin{figure}[!h]
\centering
\begin{subfigure}{\linewidth}
\centering
\begin{tabular}{@{}ccc@{}}
$\Delta_0 = 1$  &  $\Delta_0 = 3$ &  $\Delta_0 = 20$ \\
  \includegraphics[height=40mm, trim=5 5 0 0, clip]{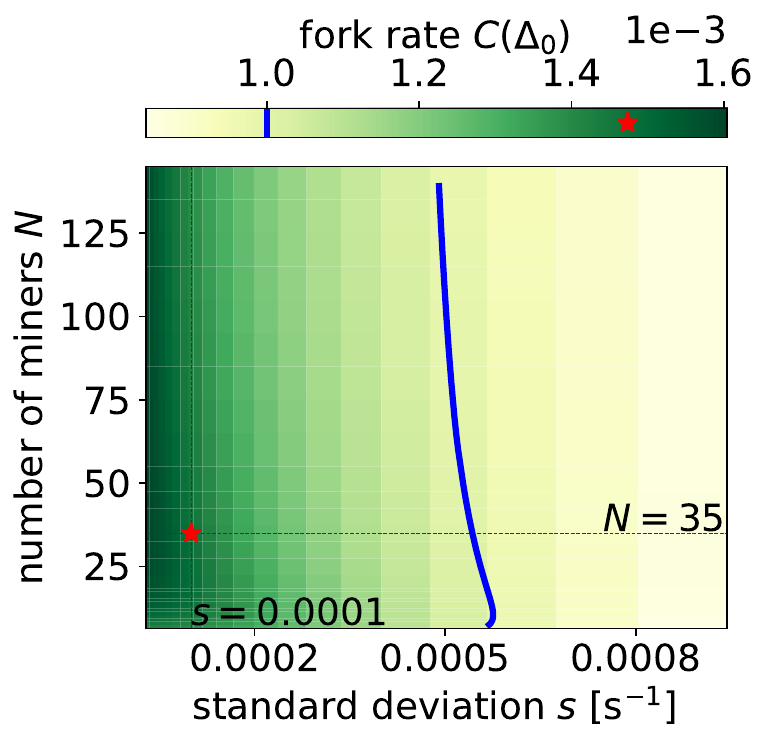}
  & 
  \includegraphics[height=40mm, trim=69 5 0 0, clip]{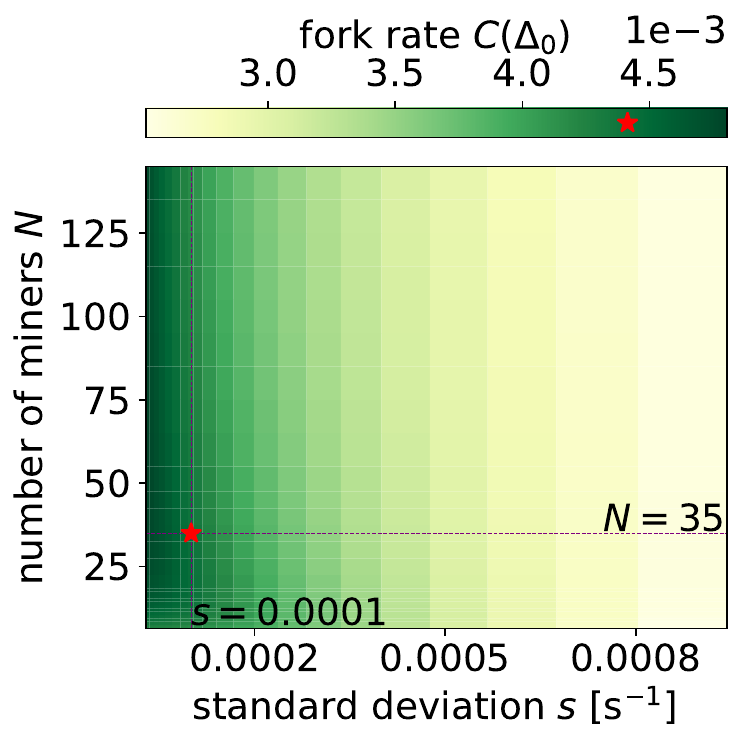}
  &  
\includegraphics[height=40mm, trim=69 5 0 0, clip]{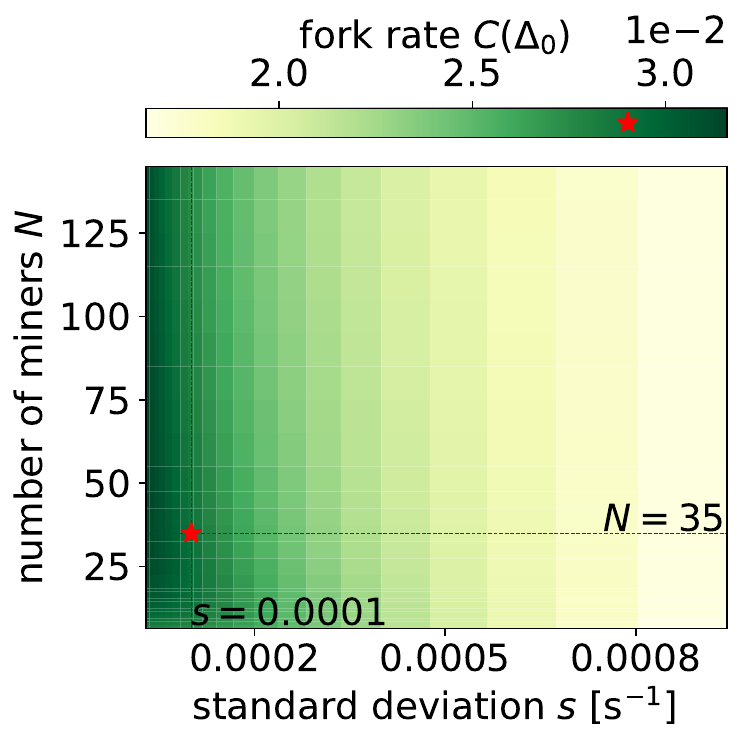}
    \end{tabular}
    \vskip -0.3cm
    \caption{Log-normal}
    
\begin{tabular}{@{}ccc@{}}
  \includegraphics[height=40mm, trim=5 5 0 0, clip]{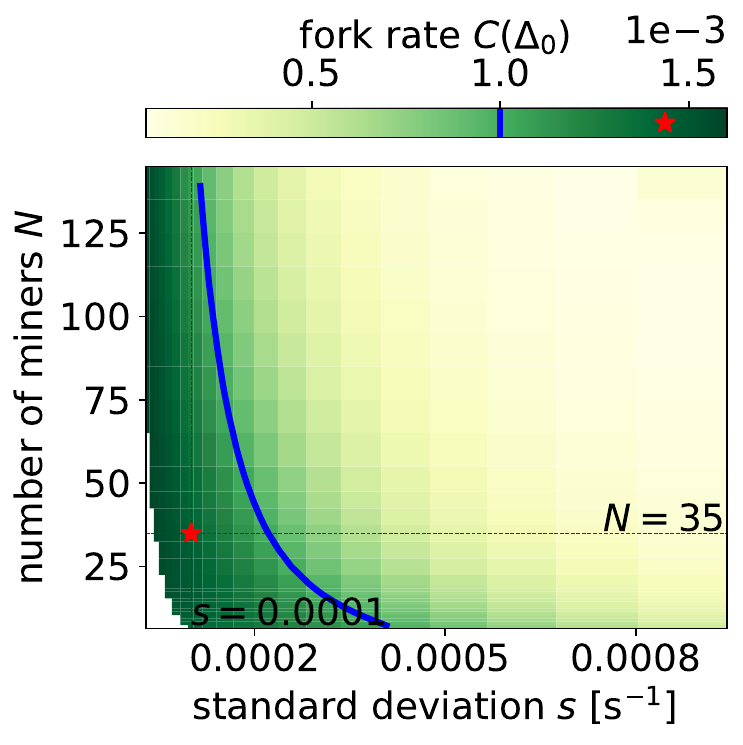}
  & 
  \includegraphics[height=40mm, trim=69 5 0 0, clip]{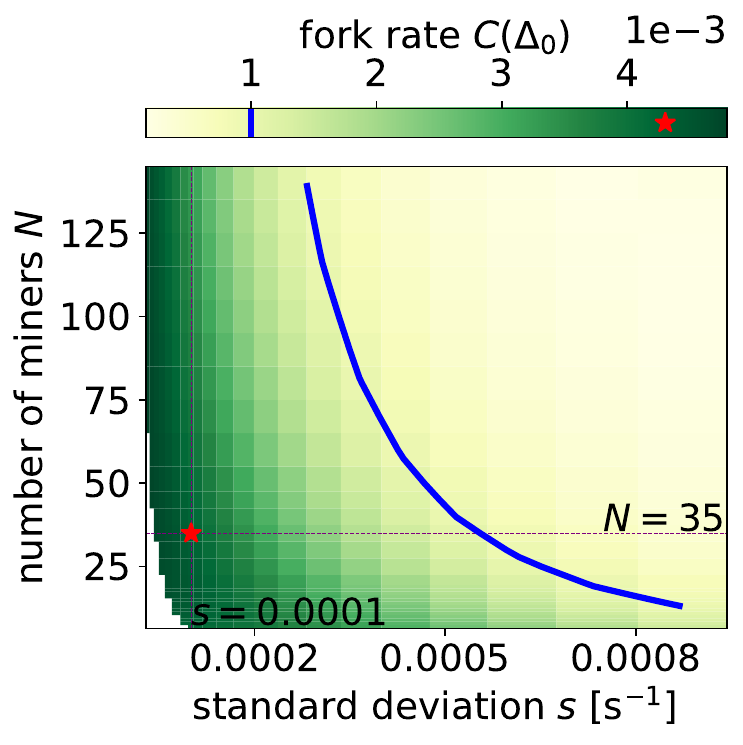}
  &  
\includegraphics[height=40mm, trim=69 5 0 0, clip]{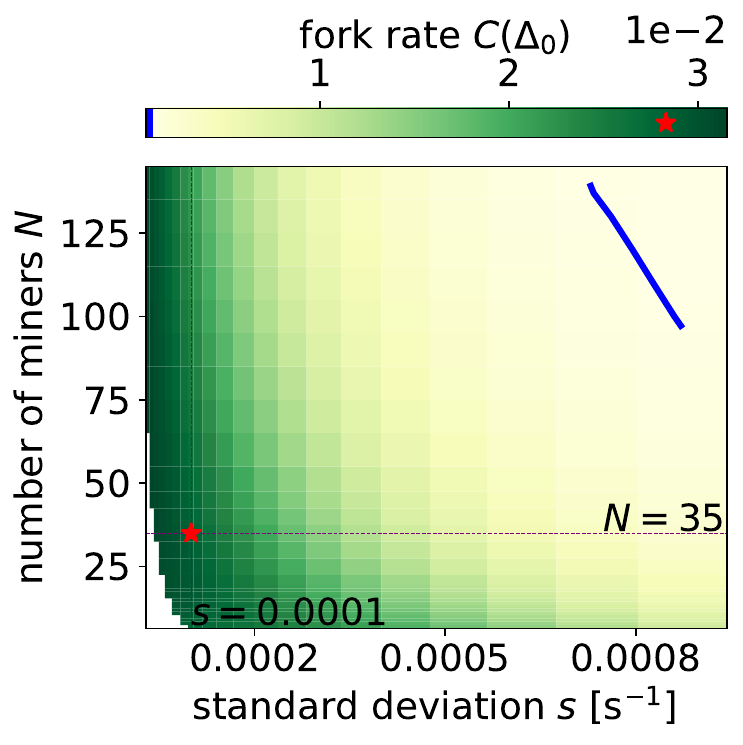}
    \end{tabular}
    \vskip -0.3cm
    \caption{Truncated power law}
\end{subfigure}
\caption{Extrapolation of fork rates under the log-normal or truncated power law hash rate distributions. Each subplot shows the fork rates with different miner numbers and hash rate standard deviation, at a given block propagation delay. The red star marks model-estimated fork rate at $N=35$ and $s=0.0001$;
the blue isoline represents a fork rate of 0.1\%.
}
\label{fig:hetero}
\end{figure}


Finally, we examine how hash rates distribution heterogeneity, proxied by the standard deviation $s$, and the number of active miners $N$ would affect fork rates.
In \autoref{fig:hetero}, we show how fork rates change at different $s$ under log-normal and truncated power law distribution of hash rates, varying the number of miners $N$, for a given total hash rate $\Lambda= 0.0017s^{-1}$ across different block propagation times.
We see that the fork rate diminishes as $s$ increases, that is, when the hash power is more concentrated and heterogeneous. 
The number of miners $N$ has a similar effect: the fork rate decreases as $N$ increases, keeping $s$ constant. 
This effect is more apparent under the truncated power law than under the log-normal distribution.
The blue curve in \autoref{fig:hetero} represents an isoline for a fork rate of 0.01\%, with the red star marking the scenario of the last sample period with $N=35$ and the $s=1\times 10^{-4}$. 
The analytically derived fork rate is close to the recently observed 0.1\% at $\Delta_0=1$ but is above it when $\Delta_0$ is higher. 

In summary, \autoref{fig:hetero} shows the offsetting effects on fork rate from a decreasing number of miners $N$ and increasing miner hash rate $s$. This explains why in \autoref{fig:forkrate_time_series} it appears that the block propagation time $\Delta_0$ is the only driver on model-estimated fork rates-- it is because historically $N$ and $s$ have been moving in opposite directions (see \autoref{subtab:hash_empirical} in the Supplementary Material) and their respective effects on estimated fork rates have been canceling out.

\section{Discussion}
We designed a mathematical model that can effectively characterize the impact of hash rate heterogeneity and network delay on the occurrence of soft forks in PoW blockchains. We considered a set of heterogeneous miners, and computed analytically the probability of observing a given time difference between the two smallest mining times by averaging over the distribution of hash rates between miners.  
On one hand, the model fits nicely to the empirical observations and the predicted fork rates are comparable with the empirical stale blocks rate. On the other hand, where the two values differ, i.e. the fraction of forks is lower or higher than the modeled value, it provides an indication that some fundamental conditions of the protocol might be breached, e.g. miners not acting independently or miners having faster communication than most other nodes in the Bitcoin network.   
The model provides a fundamental relationship which allows the fast estimation and calibration of natural fork rates in blockchain systems, for arbitrary distributions of hash rates.
Further work will be devoted to integrate in this framework the modeling of the propagation time, and as a consequence, the modeling of the network propagation process together with the mining process. This will allow not only to flag discrepancies between the real and theoretical fork rates that should emerge based on the distributions of hash rates in a set of independent miners, but also to associate these discrepancies to specific heterogeneous network properties of the miners.

The model takes as input the hash rate distribution, number of active miners and expected network delay in block propagation to calculate the probability that a soft fork occurs. We explore several specifications of hash rate distributions, considering both ex-ante distributions fitted from empirical mining data and a non-parametric Bayesian empirical distribution. 

Our study is not exempt from limitations. The most important ones are the lack of detail of intra-pool dynamics, the assumption that hash rates are constant within a 20,000 blocks time period and the relatively strong assumption that miners start mining at the same time. While these assumptions were necessary to get to our mathematical description, it is possible that at least some of them could be relaxed and we encourage further research on the topic.

Despite the limitations, we find a good agreement between the predicted fork rates, and in particular they are consistent with the empirical frequency of soft forks when we consider as propagation delay the time it takes for a block to reach 50\% of nodes in the peer-to-peer network. This would suggest that Bitcoin miners are operating in fair competition, but that most of them have better connectivity than the majority of nodes in the network. Note however that, as most of the miners we consider are actually mining pools, the results would also suggest that cooperation between miners might still be occurring within pools, either through co-location to reduce network delay or by other forms of coordination, as the empirical fork rate is consistent with the model assumption of no intra-pool latency and competition.
Beyond its empirical validation, the model formally describes the positive influence of block propagation delay and total mining power over fork rate. It also suggests that a stronger concentration of hash rates---reflected by the fat-tailedness of the distribution---reduces the expected fork rate. 

Moreover, we confirm a tendency towards hash rate concentration over time (as shown in \autoref{tab:dists}) which is a potential source of concern. Indeed while this would imply lower fork rates and a more efficient consensus mechanism, it introduces security risks that are not modeled in our framework and that could make the network more vulnerable. Further research is needed to fully explore this trade-off, and we envision our results and methodology to provide a robust mathematical setting for future works in this direction.

\section{Methods}
\label{sec:data}

\subsection{Data collection and pre-processing}

The mining dynamic in the Bitcoin network naturally changes as miners join and abandon the mining activity.
We thus divide our data chronologically, and consider a period of 20,000 blocks ($\sim$4.5 months) to be approximately stationary. The desired statistics on blocks and miners are then computed period-wise (\autoref{tab:dists}). 

\subsubsection{Block propagation delays}

We obtain propagation times from the Decentralized Systems and Network Services Research Group at KASTEL\footnote{\url{https://www.dsn.kastel.kit.edu/bitcoin/data/invstat.gpd}, accessed 2024-10-13.}---a reputable source often used by related academic works (e.g. \cite{Neudecker2019}), which report the empirical block propagation times to reach 50\%, 90\% and 99\% of the Bitcoin network approximately every hour starting from July 2015. 

\subsubsection{Forks} The public GitHub repository \texttt{bitcoin-data/stale-blocks} contains up-to-date, crowd-sourced information on stale blocks and their respective block number. We complement this dataset with anecdotal fork data published by Neudecker et al. \cite{Neudecker2019} and orphan block data from Blockchain.com. 
We then compute the fork rate at a given period as the number of block heights at which at least 1 stale or orphan block appeared, divided by the total number of blocks in that period.
The \texttt{bitcoin-data/stale-blocks} is likely to under-report stale blocks due to its crowd-sourced nature: to reconcile it with other data sources, we adjust the computed fork rate to match the value of 0.41\% reported by Gervais et al. \cite{Gervais2016OnBlockchains} at the end of February 2016. This results in a rescaling of the fork rates by a factor of 1.476. 

\subsubsection{Miners}

In our framework, a miner can be an independent miner or can be a mining pool. In case of the latter, we assume that there is no propagation delay among the members within the same mining pool. To obtain empirical miner information, we fetch coinbase transaction information including \texttt{block\_time}, \texttt{block\_timestamp} and \texttt{output.addresses} of all the blocks in the Bitcoin blockchain from the public dataset \texttt{crypto\_bitcoin} on Google BigQuery. We then identify blocks mined by mining pools by looking up coinbase trasactions' output addresses on the mining pool dictionary data from \hyperlink{https://www.blockchain.com/explorer/charts/pools}{Blockchain.com} and by looking up the block number on the miner data for selected blocks by Neudecker et al. \cite{Neudecker2019}\footnote{\url{https://www.dsn.kastel.kit.edu/bitcoin/forks/forks.gpd} and \hyperlink{https://explorer.cloverpool.com/btc/blocks}{cloverpool.com}, accessed 2024-10-15.}. For the remaining unidentified mining addresses, we use a simple heuristic to cluster: if two addresses ever appear in the same coinbase transaction, we consider them to belong to the same miner (which can be a mining pool). The total number of miners $N$ is thus estimated as the number of those mining clusters.
\autoref{fig:miner_stacked} illustrate the distribution of blocks mined among miners throughout time.

\subsubsection{Hash rates}
We retrieve the daily average total hash rate, that is, the average number of hashes per second that can be generated by the aggregate network on a specific day, on the Bitcoin network through the glassnode API\footnote{\url{https://api.glassnode.com/v1/metrics/mining/hash_rate_mean?a=BTC&i=24h}, accessed 2024-10-13.}.
We additionally convert the value of \texttt{bits} of each block fetched from \texttt{crypto\_bitcoin} dataset to the level of mining difficulty (see \autoref{lst:bitcoin_difficulty}); the resultant difficulty value describes the expected number of hashes needed to mine one block. 
For each observation period, the normalized total hash rate $\Lambda$ (in blocks/second) is thus computed as the mean ratio between total hash rate (in hashes/second) and difficulty (in hashes/block). For each period, the calculated $\Lambda$ value equals approximately the reciprocal of block time $t_\text{min}$ (see \autoref{subtab:hash_empirical}), which is in accordance with our model description (Eq. \ref{eq:block_time}).

%

\subsubsection{Hash rate distribution estimation}

For each period, we count the number of blocks $\{b_i\}$ mined by each distinct miner $i \in \{1,2,...,N\}$, where we recall $\sum_i^N b_i = B = 20,000$. Miners' hash rates $\{\lambda_i\}$ are then estimated as $\{\tfrac{b_i \Lambda}{B}\}$. 
We can then apply the method of moments to estimate the parameters the null distributions presented in \autoref{sec:null} from the empirical mean $m = \tfrac{\Lambda}{N}$ and standard deviation $s = \sqrt{\mathrm{Var}\tfrac{b_i\Lambda}{B}}$; specifically, the rate parameter $r$ of $\text{Exp}(r)$,
$\mu$ and $\sigma$ of $\text{LN}(\mu, \sigma^2)$, 
as well as 
$\alpha$ and $\beta$ of truncated power law $\text{TPL}(\alpha, \beta)$ (see \autoref{sec:null}) are estimated as follows:%
\begin{align}
r = \tfrac{1}{m}
\quad
\sigma = \sqrt{\ln{\left[
1+(\tfrac{s}{m})^2
\right]}}
\quad
\mu = \ln m - \tfrac{s^2}{2}
\quad
\alpha = 1 - \left(\tfrac{m}{s}\right)^2
\quad
\beta = \tfrac{m}{s^2}
\label{eq:mom}
\end{align}%
We display the estimated parameters in  \autoref{subtab:hash_dis}.

\printbibliography

\clearpage

\appendix

\part*{Supplementary Material}

\section{Formula derivation}

Let $CC(\Delta) = \int_D p(\{\lambda_i\})  
    \left( \frac{\sum_i^N \frac{ \lambda_i}{e^{\Delta \sum_{j \neq i} \lambda_j}}}{\sum_i^N\lambda_i}\right) d\{\lambda_i\}$ denotes the complementary cumulative function representing the probability of no fork. We now consider the convenient integral expression for the denominator:
%
$\frac{1}{\sum_i^N\lambda_i} = \int_{0}^{\infty}e^{-x\sum_i^N \lambda_i}dx
$.
%
Thus, 
\begin{align*}
CC(\Delta) &= \int_D p(\{\lambda_i\})  
    \left(
    \int_{0}^{\infty}e^{-x\sum_i^N \lambda_i}dx 
    \sum_i^N \frac{ \lambda_i}{e^{\Delta \sum_{j \neq i} \lambda_j}}
    \right)
    \\ &= \int_{0}^{\infty} dx \int_D p(\{\lambda_i\})  
    \left(
    \sum_i^N \frac{ \lambda_i}{e^{\Delta \sum_{j \neq i} \lambda_j + x\sum_i^N \lambda_i}}
    \right) 
    \\ &= \int_{0}^{\infty} dx 
    \underbrace{\int_D p(\{\lambda_i\})  
    \left(
    \sum_i^N \frac{\lambda_i}{e^{x\lambda_i}} 
    \prod_{j \neq i} \frac{1}{e^{(\Delta+x)\lambda_j}}
    \right)}_{\mathcal{A}}.
\end{align*}

When $\lambda_i > 0$ for all $i$ and $\{\lambda_i\}$ are independent, we can express $\mathcal{A}$ as

\begin{align*}
    \mathcal{A} = \int_0^{\infty} \dots \int_0^{\infty} 
    \left(
    \sum_i^N \frac{\lambda_i p(\lambda_i)d{\lambda_i}}{e^{x\lambda_i}} 
    \prod_{j \neq i} \frac{p(\lambda_j)d{\lambda_j}}{e^{(\Delta+x)\lambda_j}}
    \right)
     =
    \sum_i^N 
    \left(
     \int_0^{\infty}\frac{\lambda_i p(\lambda_i)d{\lambda_i}}{e^{x\lambda_i}} 
    \prod_{j \neq i}
    \int_0^{\infty} \frac{p(\lambda_j)d{\lambda_j}}{e^{(\Delta+x)\lambda_j}}
    \right),
\end{align*}
hence Eq. \eqref{eq:cdelta_id}.

\section{Algorithm to calculate Bitcoin mining difficulty}
\label{sec:algo_diffclty}
\begin{lstlisting}[caption={Python script to convert \texttt{bits} to difficulty value}, label={lst:bitcoin_difficulty}]
# Maximum target value (for difficulty 1, i.e., the easiest level)
MAX_TARGET = 0x00000000FFFF0000000000000000000000000000000000000000000000000000

def bits_to_difficulty(bits_hex_str: str) -> float:
    """
    Convert 'bits' field from Bitcoin block header to the expected number of hashes needed
    to mine a block at this difficulty.

    :param bits_hex_str: Hex string representing the 'bits' field (e.g., '1b00dc31').
    :return: Expected number of hashes needed to find a valid block.
    """

    # Extract exponent (first byte) and coefficient (next three bytes)
    exponent = int(bits_hex_str[:2], 16)
    coefficient = int(bits_hex_str[2:], 16)

    # Calculate the current target value
    target = coefficient * (256 ** (exponent - 3))

    # Difficulty is the ratio of max_target to the current target, scaled by 2^32
    return (MAX_TARGET / target) * (2**32)
\end{lstlisting}

\section{Considered distributions for $\{\lambda_i\}$}

\subsection{Empirical distribution}

Give the number of blocks mined by miner $i$ $\{ b_i \}$ in a certain set of blocks of size $B = \sum_i b_i$, if we use frequentist inference, then we can approximate $\lambda_i$ as
\begin{align}
\lambda_i = \frac{b_i \cdot \Lambda}{B}.
\end{align}
under which $\it HHI$ can be computed as 
\begin{align}
    {\it HHI} = \sum_i^N \left( \frac{b_i}{B} \right)^2
    \label{eq:emp_hhi}
\end{align}
%
and fork rate can be calculated as
\begin{align}
    C(\Delta_0 \vert \{\lambda_i = \frac{b_i \cdot \Lambda}{B}\}) 
    = 1-  \sum_i^N \frac{\frac{b_i}{B}}{e^{\Delta_0 \Lambda \left(1 - \frac{b_i}{B} \right)}}.
\label{eq:cdelta_freq}
\end{align}%

\subsection{Semi-empirical distributions}
\label{sec:semi-emp}

Under a Bayesian framework in order to, we can account for the uncertainty on the true value of the hash rates conditioned over a given measurement $\lbrace b_i \rbrace$. The likelihood $p(b \vert \lambda)$ of observing $b$ blocks being mined by a miner with hash rate $\lambda$ follows a Gaussian distribution
$
p(b|\lambda) = \tfrac{e^{-\tfrac{\left(b - \gamma \cdot \lambda\right)^2}{2 \gamma \cdot \lambda}}}{\sqrt{2\pi \gamma \cdot \lambda}}
$,
which is in principle unbounded over $b$ but effectively centered around the value $\gamma \cdot \lambda$, where $\gamma=\tfrac{B}{\Lambda}$. 
Applying Bayes theorem and assuming a flat prior, we can introduce the empirical distribution over $\lambda$ conditioned on a given number of mined blocks:%
\begin{align}
\label{eq:plamb}
p(\lambda|b) = \tfrac{e^{-\tfrac{\left(b - \gamma\cdot \lambda \right)^2}{2 \gamma\cdot \lambda}}}{\sqrt{2\pi \lambda/\gamma}}  .
\end{align}


\subsubsection{INID $\{\lambda_i\}$}

Assuming $\{\lambda_i\}$ are independent but not necessarily identically distributed (INID), we plug Eq. \eqref{eq:plamb} in \eqref{eq:cdelta_id} to obtain
\begin{align}
\label{eq:cdelta_id_emp}
C(\Delta_0) = 1-  \int_0^\infty  
    \tfrac{
    \sum_i^{N} 
    \left(
     1+b_i \cdot \sqrt{1+\tfrac{2x}{\gamma}}
         \right) e^{B\left(1-\tfrac{2(\Delta_0+x)}{\gamma}\right)+b_i\left(\sqrt{1+\tfrac{2(\Delta_0+x)}{\gamma}} - \sqrt{1+\tfrac{2x}{\gamma}}\right)}
    }{\gamma  \left(\sqrt{1+\tfrac{2x}{\gamma}}\right)^3  \left(\sqrt{1+\tfrac{2(\Delta_0+x)}{\gamma}}\right)^{N-1}} dx .
\end{align}

\subsubsection{i.i.d. $\{\lambda_i\}$}

In the even more specific case of $\lambda_i$ being i.i.d., we can introduce the semi-empirical marginal distribution for $\lambda$ by averaging $p(\lambda|\{b_i\})$ over $\{b_i\}$, finding
\begin{align}
\label{eq:plamb_iid}
p(\lambda|\{b_i\}) = \tfrac{1}{N}\sum_i^N\tfrac{e^{-\tfrac{\left(b_i - \gamma\cdot \lambda\right)^2}{2 \gamma \cdot \lambda}}}{\sqrt{2\pi \lambda/\gamma}} .
\end{align}
And so, plugging Eq. \eqref{eq:plamb_iid} in \eqref{eq:cdelta_iid}, we arrive at
\begin{align}
\label{eq:cdelta_iid_emp}
C(\Delta_0) = 1-  \int_0^\infty  
    \tfrac{
    \sum_i^N 
    \left(
     1+b_i \cdot \sqrt{1+\tfrac{2x}{\gamma}}
     \right) e^{b_i\left(1 - \sqrt{1+\tfrac{2x}{\gamma}}\right)}
    }{\gamma  \left(\sqrt{1+\tfrac{2x}{\gamma}}\right)^3} 
    \left[ 
        \tfrac{
    \sum_i^N 
    e^{b_i \left(1-\sqrt{1+\tfrac{2(\Delta_0+x)}{\gamma}}\right)}
    }{ N  \sqrt{1+\tfrac{2(\Delta_0+x)}{\gamma}}} 
    \right]^{N-1}dx .
\end{align}

\subsection{Null distributions for i.i.d. $\{\lambda_i\}$}\label{sec:null}
As an alternative approach to the Bayesian setting described above, we also consider 3 commonly used null distributions for i.i.d. $\{\lambda_i\}$, assuming they can be distributed according to an exponential, log-normal or truncated power law distribution. 

\subsubsection{Exponential}
We select an exponential distribution for its parsimony, as the distribution can be characterized simply by one rate parameter $r$. For $\lambda_i \sim \text{Exp}(r) \; \forall i$, we plug the density function
\begin{align}
p(\lambda_i) = r e^{-r \lambda_i}
\end{align}
into Eq. \eqref{eq:cdelta_iid} and get
\begin{align}
C(\Delta_0) 
& = 1-  N \int_0^\infty  
    \left[\left(
     \int_0^{\infty}\tfrac{\lambda rd{\lambda}}{e^{x\lambda + r \lambda} }\right) 
    \left(
    \int_0^{\infty} \tfrac{rd{\lambda}}{e^{(\Delta_0+x)\lambda+ r \lambda}}
    \right)^{N-1}
    \right] dx
    \\ & 
    = 
    1-  N r^N \int_0^\infty  \tfrac{dx}{(r+x)^2 
    (\Delta_0+r+x)^{N-1}} .
    \label{eg:c_ex}
\end{align}   

\subsubsection{Log-normal} 
A log-normal distribution is frequently observed in systems characterised by unbounded multiplicative (or proportional) growth \cite{Mitzenmacher2004ADistributions}, and there is a large body of literature supporting the view that blockchain consensus protocols are strongly affected by this accumulation dynamics \cite{Gao2024Heterogeneity-Protocol,Makarov2021BlockchainMarket,Campajola2022ThePlatforms,Campajola2022MicroVelocity:Currencies}.
For $\lambda_i \sim \text{LN}(\mu, \sigma^2) \; \forall i$, we integrate numerically after plugging its density function
\begin{align}
p(\lambda_i) = \frac{e^\frac{-(\ln \lambda_i - \mu)^2}{2 \sigma^2}}{\lambda_i \sigma \sqrt{2 \pi}}
\end{align}%
into Eq. \eqref{eq:cdelta_iid}.

\subsubsection{Truncated power law} A truncated power law distribution also known as power law distribution with exponential cut-off, typically emerges in complex systems where proportional growth effects get capped by some natural scale or limit \cite{Burroughs2001Upper-truncatedSystems}. In the specific context of mining, the cap we would see is given by the fact that the total hashrate sums to $\Lambda$, meaning that each individual miner cannot have a hashrate larger than that in a given sample.
This limitation results in a truncation of the power law at the upper end, which is generally captured by multiplying the power law function with a decreasing exponential.
As above, for $\lambda_i \sim \text{TPL}(\alpha, \beta) \; \forall i$, we can plug the density function 
\begin{align}
p(\lambda_i) =  \tfrac{\beta^{1-\alpha}}{\Gamma(1-\alpha) \cdot \lambda_i^{\alpha} \cdot e^{\beta \lambda_i}}
\end{align}%
into Eq. \eqref{eq:cdelta_iid} and arrive at
\begin{align}
C(\Delta_0) 
& = 1-  N (1-\alpha) \cdot \beta^{N (1-\alpha)}  \int_0^\infty  
    \frac{dx}{
    (\beta + x)^{2-\alpha} \cdot 
    (\Delta_0+ \beta + x)^{(N-1)(1-\alpha)}
    }.
\label{eg:c_tpl}
\end{align}   
For succinctness we omit the derivation steps. Note that the truncated power law distribution can be deemed as a generalization of the exponential distribution. Specifically, when $\beta = r$ and $\alpha = 0$, $\text{TPL}(\alpha, \beta)$ can be reduced to $\text{Exp}(r)$.




\begin{figure}[bt]
\begin{subfigure}{0.32\linewidth}
\includegraphics[height=0.18\textheight,trim=20 0 145 20, clip]{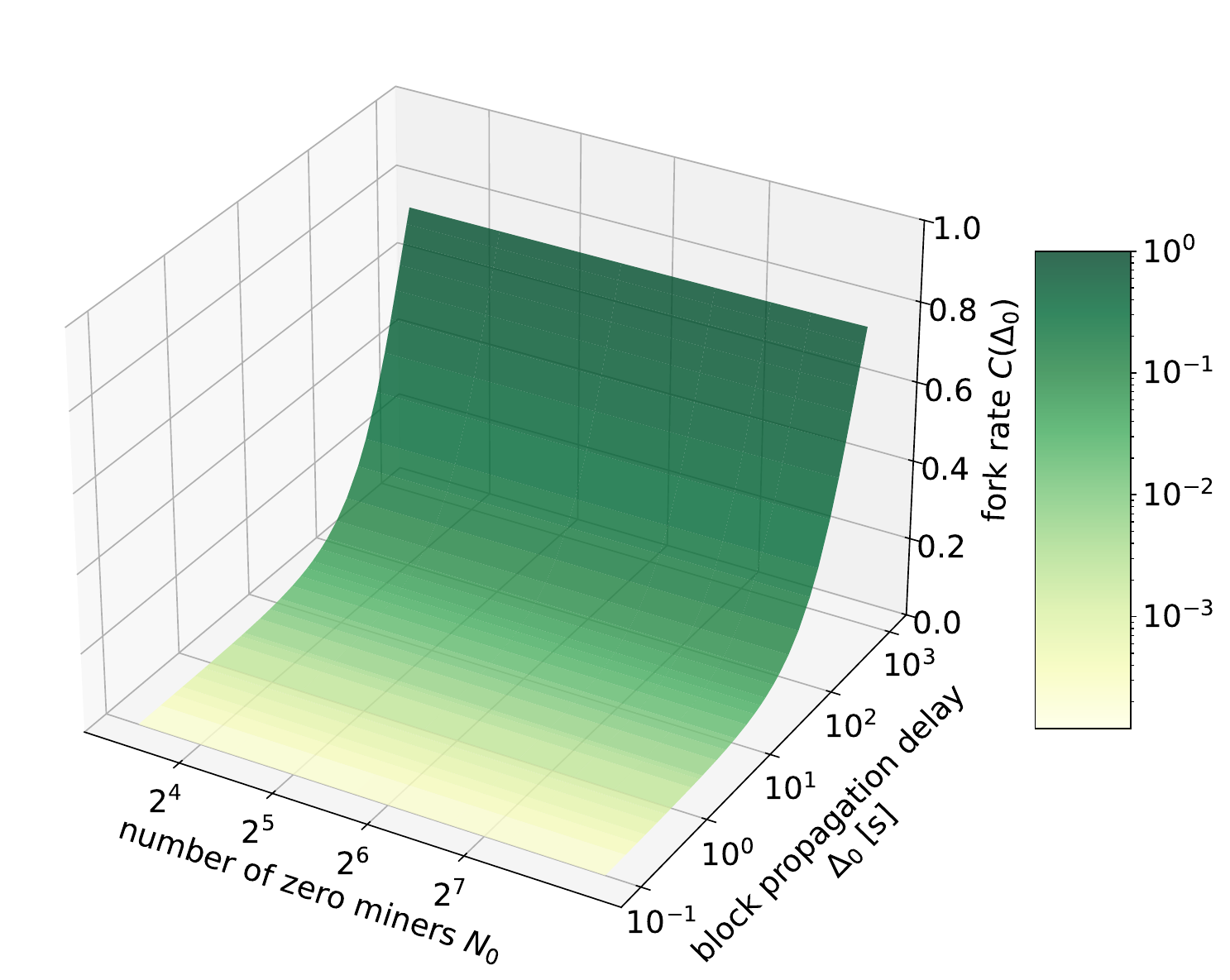}
\vskip -0.2cm
\caption{Semi-emp, independent}
\label{fig:surface_emp_id}
\end{subfigure}
\begin{subfigure}{0.32\linewidth}
\includegraphics[height=0.18\textheight,trim=20 0 10 20, clip]{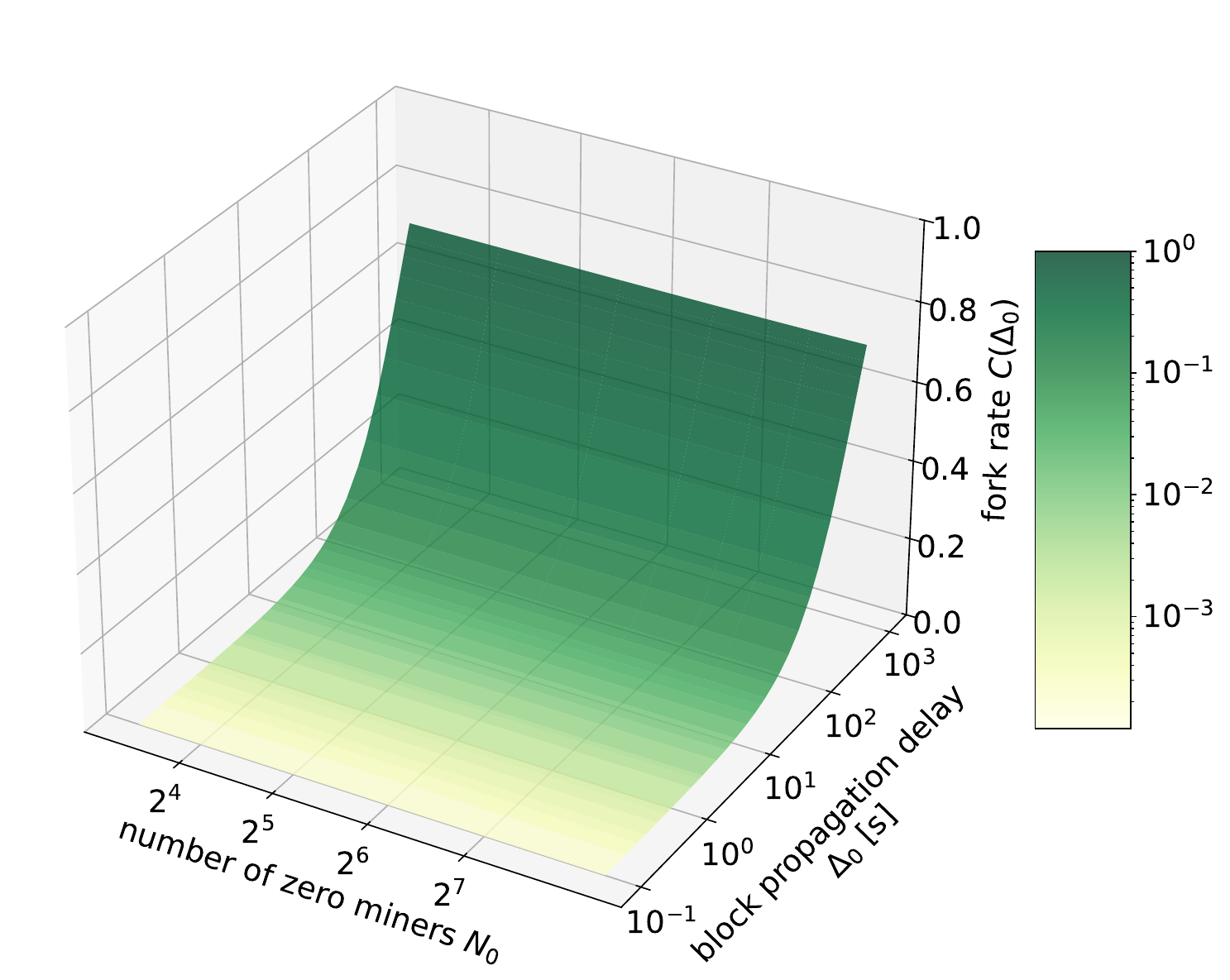}
\vskip -0.2cm
\caption{Semi-emp, i.i.d.}
\label{fig:surface_emp_iid}
\end{subfigure}
\\
\begin{subfigure}{0.32\linewidth}
\includegraphics[height=0.18\textheight,trim=20 0 145 20, clip]{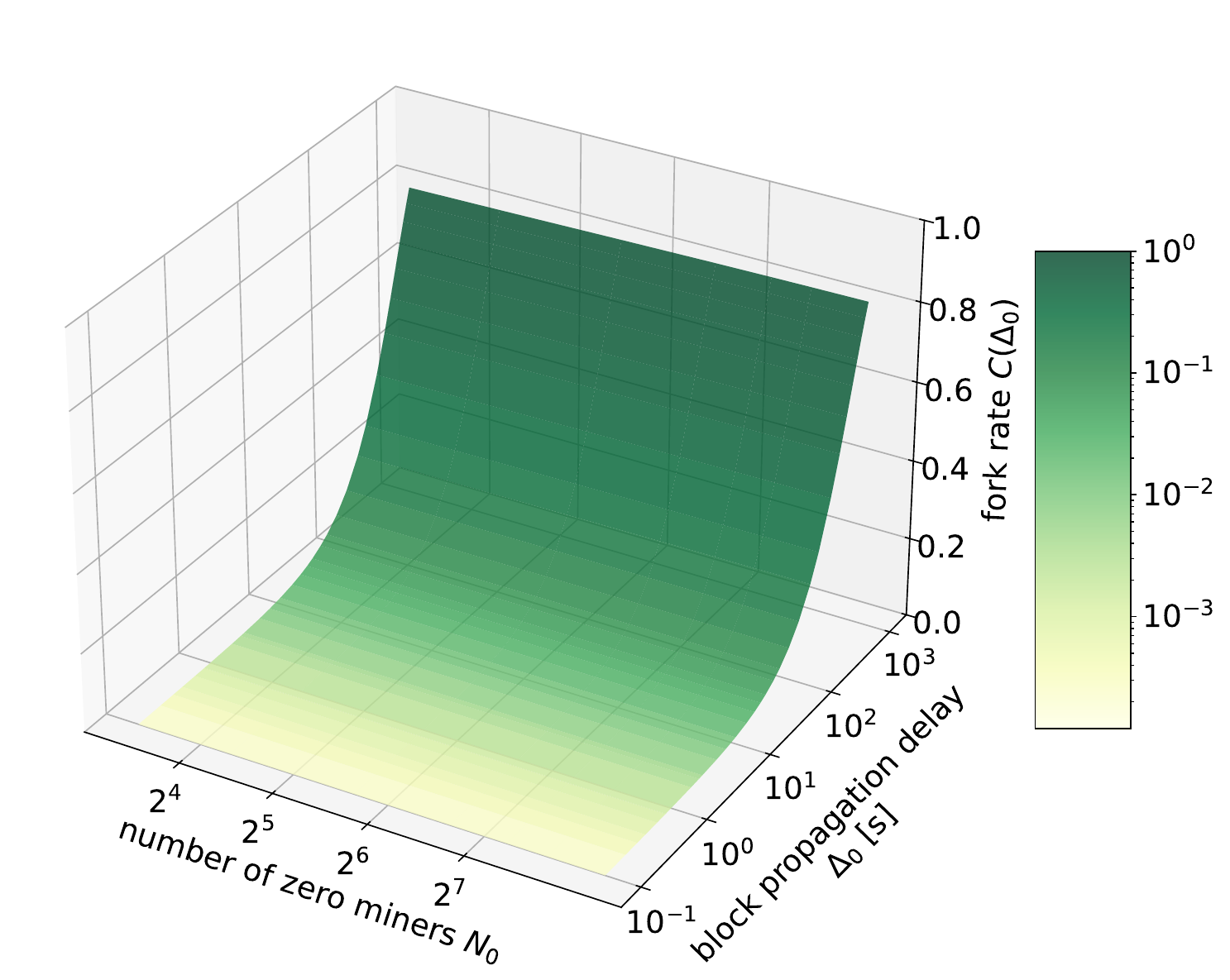}
\vskip -0.2cm
\caption{Exponential}
\label{fig:surface_exp}
\end{subfigure}
\begin{subfigure}{0.32\linewidth}
\includegraphics[height=0.18\textheight,trim=20 0 145 20, clip]
{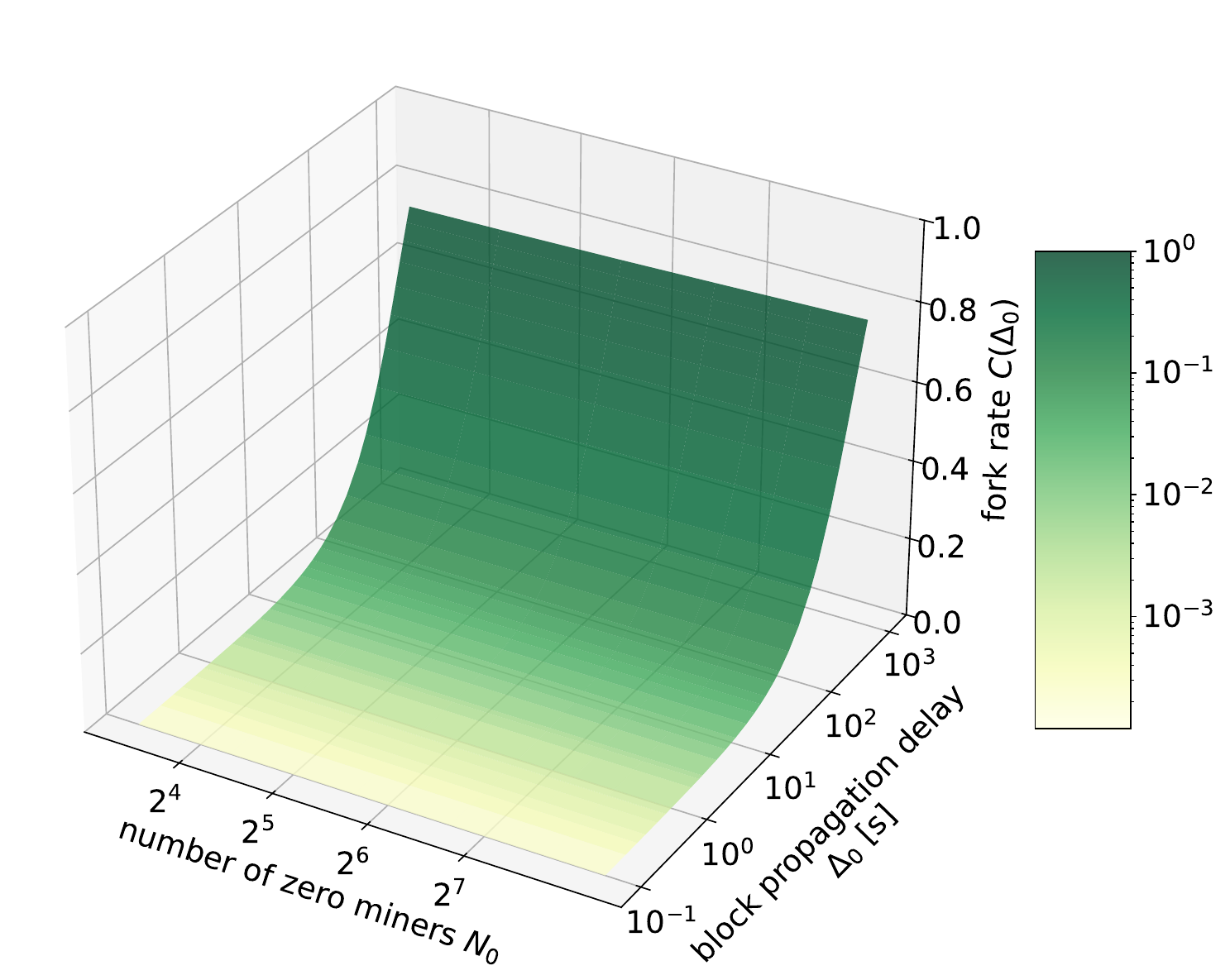}
\vskip -0.2cm
\caption{Log normal}
\label{fig:surface_lognormal}
\end{subfigure}
\begin{subfigure}{0.34\linewidth}
\includegraphics[height=0.18\textheight,trim=20 0 10 20, clip]{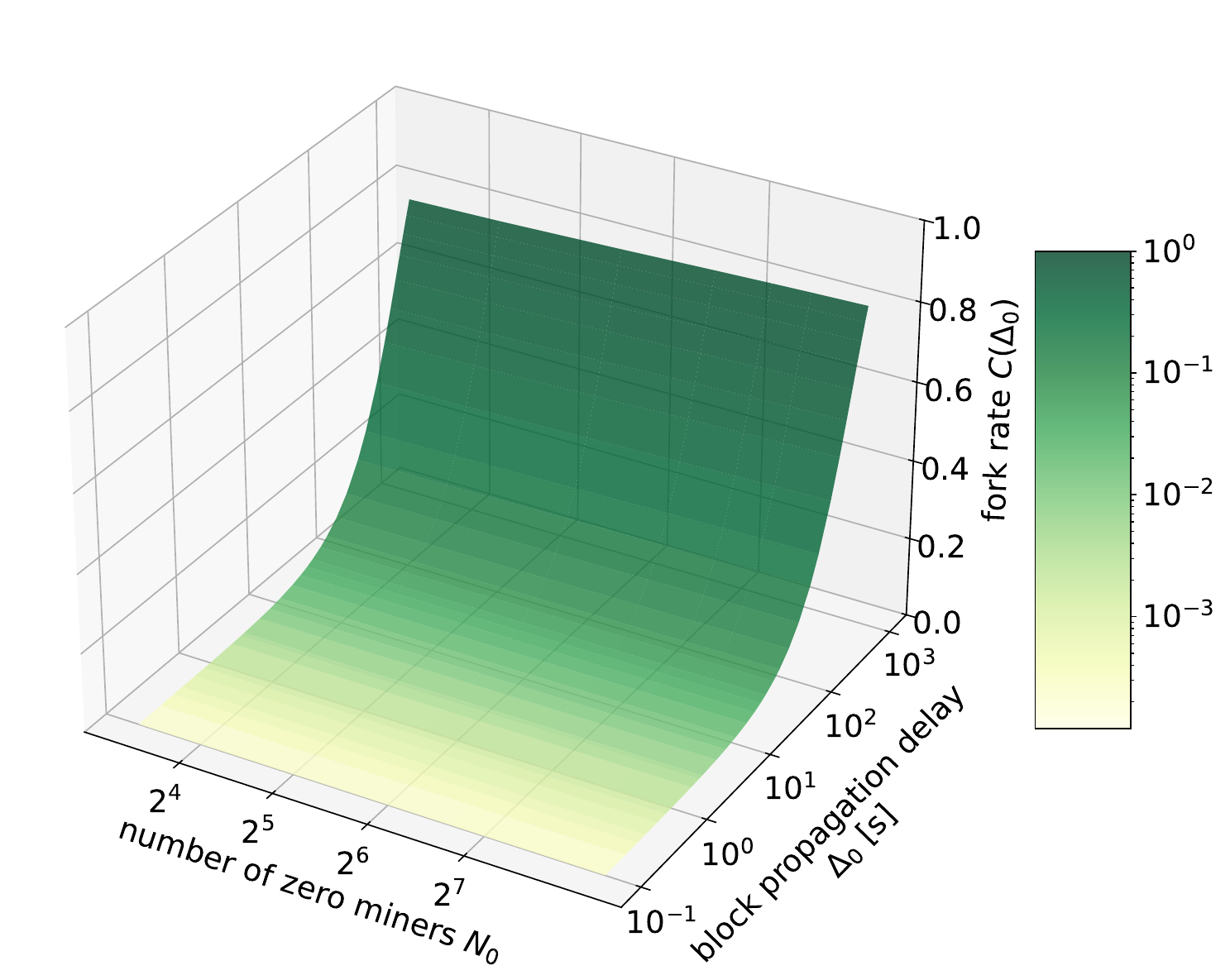}
\vskip -0.2cm
\caption{Truncated power law}
\label{fig:surface_lomax}
\end{subfigure}
\caption{Fork rates by number of zero miners and block propagation delay at various given hash rate distributions.}
\label{fig:surface}
\end{figure}

\subsubsection{Different number of \enquote{zero miners}}
\label{sec:zero}

\begin{figure}[t]
    \begin{subfigure}[t]{0.328\textwidth}
    \includegraphics[height=0.17\textheight, trim=0 8 0 0, clip]{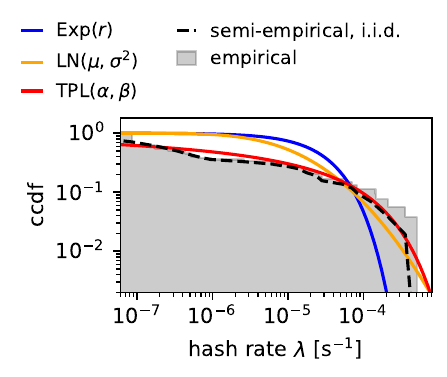}
    \vskip -0.2cm
    \caption{20 zero miners}
    \end{subfigure}
    \hfill
    \begin{subfigure}[t]{0.328\textwidth}
    \includegraphics[height=0.17\textheight, trim=0 8 0 0, clip]{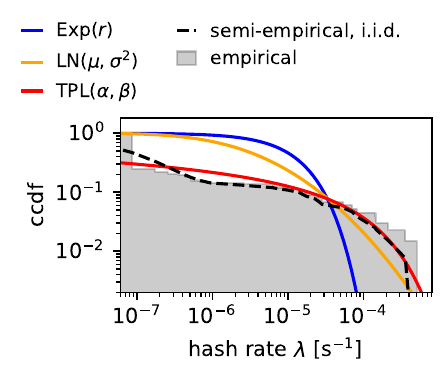}
    \vskip -0.2cm
    \caption{100 zero miners}
    \end{subfigure}
    \hfill
    \begin{subfigure}[t]{0.328\textwidth}
    \includegraphics[height=0.17\textheight, trim=0 8 0 0, clip]{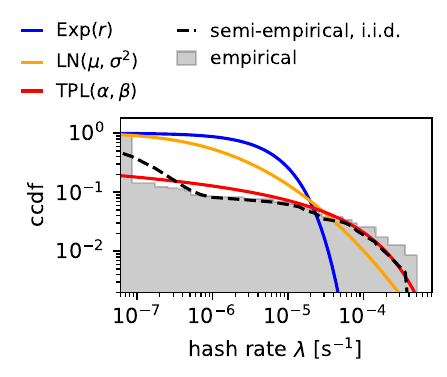}
    \vskip -0.2cm
    \caption{200 zero miners}
    \end{subfigure}
    
\caption{Complementary cumulative distribution function (ccdf) of hash rates when different numbers of \enquote{zero miners} are added to observed miners.}
\label{fig:zerodist}
\end{figure}

When estimating hash rates from mined block frequencies, we may need to consider the possibility of existent miners that have not been able to mine a block in the period, who we term as \enquote{zero miners}.
In general, assuming a fair sampling, the probability of mining should be simply proportional to the hash rate and the inability of a miner to mine in a long time horizon should reflect a very low hash rate. Nevertheless it is possible, though improbable, that there exists a number of unlucky miners that did not manage to mine a block in a relatively long fraction of time.
In \autoref{fig:zerodist}, we depict the scenarios of different numbers of zero miners being added to the miners observed from our last sample period (\autoref{fig:lastperiod}). For a given number of zero miners, we refit the distributions under the i.i.d. assumption. 
The fork rates calculated with the refitted distribution with different numbers of zero miners $N_0$ are depicted in \autoref{fig:surface_emp_iid}-\ref{fig:surface_lomax}.

We additionally check the scenario with miner hash rates following independent, but non-identical distributions applying the semi-empirical approach.
%
The distribution of a zero miner's hash rate is essentially
the posterior distribution conditioned on zero blocks, which can be written out by plugging $b=0$ into Eq. \eqref{eq:plamb}:
$
p(\lambda|0) = \tfrac{e^{-\tfrac{\gamma \cdot \lambda}{2}}}{\sqrt{2\pi \lambda/\gamma}}
$. The corresponding result is illustrated in \autoref{fig:surface_emp_id}. Overall, the effect of \enquote{zero miners} is almost negligible regardless of the distribution type used or block propagation delay.






\section{Simulations}

To validate our analytical derivations, we conduct simulations within a synthetic blockchain environment.
For each round, we initiate the environment with $N \geq 2$ miners, whose hash powers $\{\lambda_i\}_{i = 1, 2, ..., N}$ are randomly sampled from a null distribution, 
To simulate the mining process, we generate for each miner $i$ a mining time $t_i$ following an exponential distribution with rate $\lambda_i$, i.e., $t_i \sim \text{Exp}(\lambda_i)$, mimicking the Poisson arrival pattern of mined blocks (see \autoref{sec:model}). We calculate the time difference $\Delta$ between the two fastest miners for each round. A fork is deemed to have occurred if $\Delta$ is below the pre-specified block propagation time $\Delta_0$; otherwise, the round concludes with no fork.
%
For a given set of hyper-parameters, i.e. the number of miners $N$, mean $m$ and standard deviation $s$ of the hash power distribution, and $\Delta_0$, we iterate for $n$ rounds---where $n$ is sufficiently large ($n=10^7$ in our experiments)---and count the number of rounds with a fork denoted as $n_\text{fork}$. The simulated fork rate is then calculated as the fraction of simulations where a fork occurred, $\frac{n_\text{fork}}{n}$.

\begin{figure}[t]
\centering
    \begin{subfigure}[t]{0.3\textwidth}
    \includegraphics[width=\linewidth,trim=20 20 20 0, clip]{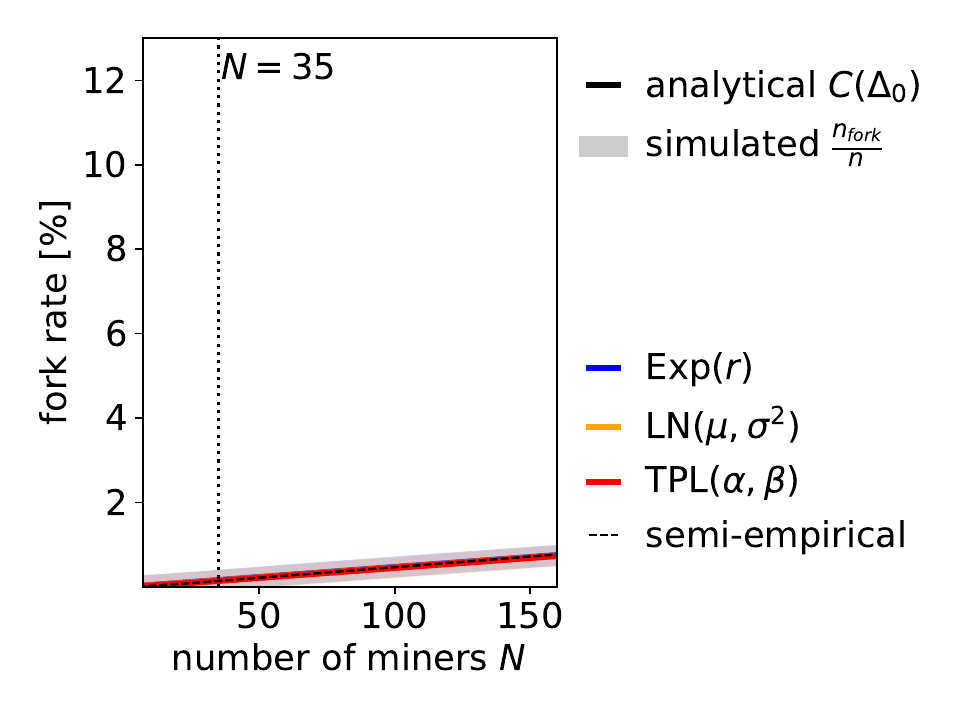}
    \caption{$\Delta_0 = 1$ [s]}
    \end{subfigure}
    \hfill
    \begin{subfigure}[t]{0.3\textwidth}
    \includegraphics[width=\linewidth,trim=20 20 20 0, clip]{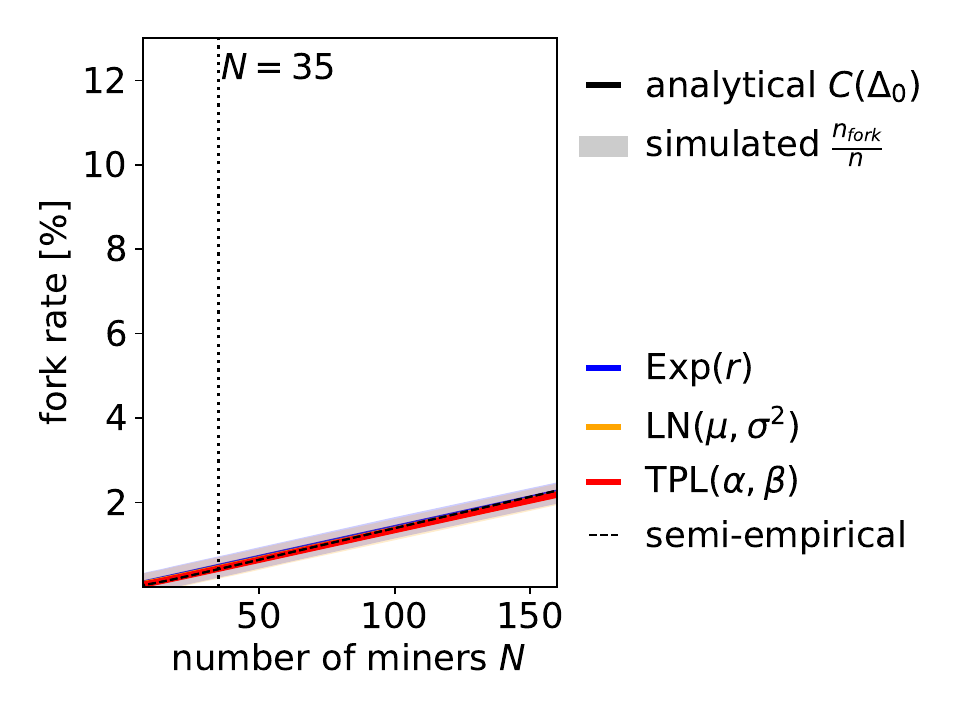}
    \caption{$\Delta_0 = 3$ [s]}
    \end{subfigure}
    \hfill
    \begin{subfigure}[t]{0.3\textwidth}
    \includegraphics[width=\linewidth,trim=20 20 20 0, clip]{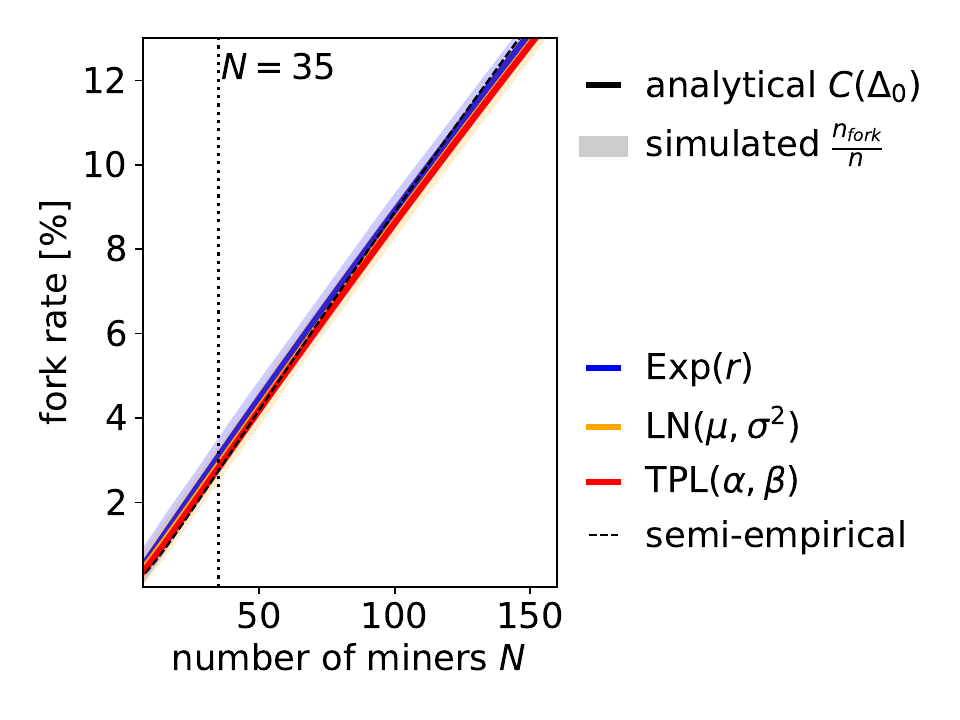}
    \caption{$\Delta_0 = 20$ [s]}
    \end{subfigure}
\caption{Fork rates computed from the simulated environment (thick, transparent curves) compared with those calculated analytically (thin, opaque curves) under various block propagation delays, hash rate distributions, and miner numbers.} 
\label{fig:forkrate_per_delta}
\end{figure}

In \autoref{fig:forkrate_per_delta}, we superimpose the analytical results  (thin, opaque curves) on the simulated output (thick, transparent curves) for the null distributions---exponential $\text{Exp}(r)$, log-normal $\text{LN}(\mu,\sigma^2)$ and truncated power law $\text{TPL}(\alpha, \beta)$. The distribution parameters are calculated as Eq.~\eqref{eq:mom} at $m=5\times 10^{-5}$ and $s=1\times10^{-4}$, approximately the same level of the mean and standard deviation of hash rates estimated from our last few sample periods (see \autoref{tab:dists}). We also compute the fork rate using the semi-empirical method under the i.i.d. assumption. Specifically, we assume each miner's hash rate is randomly sampled from the same distribution with the PDF estimated following Eq. \eqref{eq:plamb_iid} using the numbers of blocks mined from the 35 miners and the total hash rate $\Lambda = 0.0017$ [s$^{-1}$]. The results are plotted in dashed red curves.

\autoref{fig:forkrate_per_delta} shows that our analytical result correctly predicts the simulation with remarkable accuracy regardless of the underlying distribution, and that, holding hash rate mean $m$ and standard deviation $s$ constant, 
the fork rate increases with the number of participating miners and the block propagation delay, as could be expected. 

\begin{landscape}
\begin{table}[pbt]
\tiny
\caption{Empirical and fitted distributions of hash rates for 24 consecutive, non-overlapping periods, each with a fixed-length window of 20,000 blocks.}
            \label{tab:dists}
    \centering
    \begin{subtable}{0.8\linewidth}

        \caption{Empirical data}
            \label{subtab:hash_empirical}
        \resizebox{1.1\textheight}{!}{\begin{tabular}{@{}llrrrrrrrrrrrrr@{}}
\toprule
\multicolumn{2}{c}{period of blocks}  &  \multicolumn{3}{c}{propagation time} & & &  &  \multicolumn{7}{c}{empirical miner hash rate}\\
\cmidrule(lr){1-2} \cmidrule(lr){3-5} \cmidrule(lr){9-15}
start \# & start time & 50\% & 90\% & 99\% & $\overline{\text{block time}}$ & fork rate & miners & $\sum(\text{hash rate})$ & mean & std & skewness & kurtosis & hhi & max share\\
 &  & [s] & [s] & [s] & $t_{\text {min }}$ [s] & [\%] & $N$ & $\Lambda$ [s$^{-1}$] & $m$ [s$^{-1}$] & $s$ [s$^{-1}$] &  &  &  & [\%]\\
\cmidrule(lr){1-2} \cmidrule(lr){3-5} \cmidrule(lr){6-8} \cmidrule(lr){9-15}
360000 & 2015-06-08 & \databarred{7.01} & 16.51 & 26.61 & 586.1 & \databarblue{0.982} & \databarpurple{85} & 0.00171 & \databarorangeone{0.000020} & \databarorangetwo{0.000060} & 3.82 & 14.73 & \databarbrown{0.12} & \databarbrown{19.34} \\
380000 & 2015-10-22 & \databarred{7.11} & 18.03 & 27.73 & 546.2 & \databarblue{0.856} & \databarpurple{69} & 0.00185 & \databarorangeone{0.000027} & \databarorangetwo{0.000082} & 3.81 & 14.52 & \databarbrown{0.15} & \databarbrown{22.89} \\
400000 & 2016-02-25 & \databarred{5.87} & 15.68 & 27.24 & 583.3 & \databarblue{0.339} & \databarpurple{35} & 0.00172 & \databarorangeone{0.000049} & \databarorangetwo{0.000106} & 2.71 & 6.98 & \databarbrown{0.16} & \databarbrown{24.44} \\
420000 & 2016-07-09 & \databarred{4.09} & 12.33 & 25.67 & 584.8 & \databarblue{0.384} & \databarpurple{56} & 0.00172 & \databarorangeone{0.000031} & \databarorangetwo{0.000065} & 2.67 & 6.89 & \databarbrown{0.10} & \databarbrown{17.80} \\
440000 & 2016-11-22 & \databarred{3.11} & 10.49 & 24.24 & 566.5 & \databarblue{0.273} & \databarpurple{56} & 0.00177 & \databarorangeone{0.000032} & \databarorangetwo{0.000057} & 2.36 & 5.32 & \databarbrown{0.08} & \databarbrown{14.12} \\
460000 & 2017-04-02 & \databarred{1.96} & 10.13 & 23.63 & 565.0 & \databarblue{0.280} & \databarpurple{62} & 0.00179 & \databarorangeone{0.000029} & \databarorangetwo{0.000052} & 2.22 & 4.30 & \databarbrown{0.07} & \databarbrown{12.55} \\
480000 & 2017-08-10 & \databarred{1.09} & 7.39 & 19.98 & 560.7 & \databarblue{0.184} & \databarpurple{75} & 0.00183 & \databarorangeone{0.000024} & \databarorangetwo{0.000054} & 2.77 & 6.86 & \databarbrown{0.08} & \databarbrown{12.80} \\
500000 & 2017-12-18 & \databarred{0.54} & 4.06 & 16.25 & 555.9 & \databarblue{0.125} & \databarpurple{66} & 0.00180 & \databarorangeone{0.000027} & \databarorangetwo{0.000069} & 3.70 & 15.25 & \databarbrown{0.11} & \databarbrown{22.42} \\
520000 & 2018-04-26 & \databarred{0.41} & 2.08 & 13.33 & 569.1 & \databarblue{0.052} & \databarpurple{74} & 0.00177 & \databarorangeone{0.000024} & \databarorangetwo{0.000060} & 3.28 & 10.85 & \databarbrown{0.10} & \databarbrown{17.90} \\
540000 & 2018-09-05 & \databarred{0.47} & 2.27 & 14.77 & 613.4 & \databarblue{0.037} & \databarpurple{89} & 0.00165 & \databarorangeone{0.000018} & \databarorangetwo{0.000050} & 3.42 & 11.61 & \databarbrown{0.09} & \databarbrown{16.75} \\
560000 & 2019-01-25 & \databarred{0.57} & 3.84 & 18.10 & 586.2 & \databarblue{0.044} & \databarpurple{77} & 0.00172 & \databarorangeone{0.000022} & \databarorangetwo{0.000054} & 3.34 & 11.66 & \databarbrown{0.09} & \databarbrown{17.21} \\
580000 & 2019-06-09 & \databarred{0.46} & 3.22 & 17.83 & 566.5 & \databarblue{0.066} & \databarpurple{67} & 0.00178 & \databarorangeone{0.000027} & \databarorangetwo{0.000061} & 3.28 & 11.39 & \databarbrown{0.09} & \databarbrown{18.39} \\
600000 & 2019-10-19 & \databarred{0.40} & 2.15 & 14.41 & 590.2 & \databarblue{0.044} & \databarpurple{54} & 0.00171 & \databarorangeone{0.000032} & \databarorangetwo{0.000060} & 2.63 & 7.50 & \databarbrown{0.08} & \databarbrown{16.89} \\
620000 & 2020-03-03 & \databarred{0.49} & 3.19 & 15.78 & 599.2 & \databarblue{0.074} & \databarpurple{44} & 0.00169 & \databarorangeone{0.000038} & \databarorangetwo{0.000067} & 2.48 & 6.59 & \databarbrown{0.09} & \databarbrown{18.34} \\
640000 & 2020-07-20 & \databarred{0.60} & 3.89 & 17.15 & 595.1 & \databarblue{0.111} & \databarpurple{43} & 0.00170 & \databarorangeone{0.000039} & \databarorangetwo{0.000068} & 1.95 & 3.34 & \databarbrown{0.09} & \databarbrown{16.30} \\
660000 & 2020-12-05 & \databarred{0.75} & 4.44 & 19.22 & 593.3 & \databarblue{0.140} & \databarpurple{50} & 0.00170 & \databarorangeone{0.000034} & \databarorangetwo{0.000065} & 2.26 & 4.71 & \databarbrown{0.09} & \databarbrown{16.82} \\
680000 & 2021-04-21 & \databarred{0.46} & 2.32 & 13.70 & 616.7 & \databarblue{0.059} & \databarpurple{39} & 0.00165 & \databarorangeone{0.000042} & \databarorangetwo{0.000070} & 1.89 & 2.75 & \databarbrown{0.09} & \databarbrown{15.92} \\
700000 & 2021-09-11 & \databarred{0.38} & 1.58 & 14.13 & 579.3 & \databarblue{0.066} & \databarpurple{36} & 0.00174 & \databarorangeone{0.000048} & \databarorangetwo{0.000084} & 1.81 & 2.07 & \databarbrown{0.11} & \databarbrown{16.16} \\
720000 & 2022-01-23 & \databarred{0.40} & 1.72 & 14.32 & 591.6 & \databarblue{0.030} & \databarpurple{33} & 0.00170 & \databarorangeone{0.000052} & \databarorangetwo{0.000091} & 1.76 & 2.04 & \databarbrown{0.12} & \databarbrown{19.38} \\
740000 & 2022-06-09 & \databarred{0.34} & 1.26 & 11.30 & 589.8 & \databarblue{0.007} & \databarpurple{26} & 0.00171 & \databarorangeone{0.000066} & \databarorangetwo{0.000110} & 1.90 & 3.10 & \databarbrown{0.14} & \databarbrown{24.06} \\
760000 & 2022-10-23 & \databarred{0.43} & 1.95 & 13.28 & 590.7 & \databarblue{0.044} & \databarpurple{37} & 0.00171 & \databarorangeone{0.000046} & \databarorangetwo{0.000109} & 3.05 & 9.61 & \databarbrown{0.17} & \databarbrown{29.75} \\
780000 & 2023-03-09 & \databarred{0.86} & 4.65 & 20.23 & 590.3 & \databarblue{0.148} & \databarpurple{38} & 0.00171 & \databarorangeone{0.000045} & \databarorangetwo{0.000111} & 3.33 & 11.54 & \databarbrown{0.18} & \databarbrown{31.38} \\
800000 & 2023-07-24 & \databarred{0.87} & 4.46 & 21.92 & 584.6 & \databarblue{0.103} & \databarpurple{35} & 0.00172 & \databarorangeone{0.000049} & \databarorangetwo{0.000116} & 3.04 & 9.09 & \databarbrown{0.18} & \databarbrown{28.84} \\
820000 & 2023-12-06 & \databarred{0.89} & 4.39 & 19.81 & 585.5 & \databarblue{0.096} & \databarpurple{33} & 0.00172 & \databarorangeone{0.000052} & \databarorangetwo{0.000111} & 3.00 & 9.38 & \databarbrown{0.16} & \databarbrown{29.30} \\
\bottomrule
\end{tabular}
}
    \end{subtable}%
    \begin{subtable}{0.2\linewidth}
        \centering
        \caption{Fitted parameters}
                \label{subtab:hash_dis}
        \resizebox{0.266\textheight}{!}{\begin{tabular}{@{}rrrrr@{}}
\toprule
\multicolumn{5}{c}{fitted distributions}\\
\cmidrule(lr){1-5}
$\text{Exp}(r)$ & \multicolumn{2}{c}{$\text{LN}(\mu, \sigma^2)$} & \multicolumn{2}{c}{$\text{TPL}(\alpha, \beta)$}\\
\cmidrule(lr){1-5}
$r$ & $\mu$ & $\sigma$ & $\alpha$ & $\beta$\\
\cmidrule(lr){1-1} \cmidrule(lr){2-3} \cmidrule(lr){4-5}
49,565 & -11.96 & 1.52 & 0.89 & 5,518 \\
37,247 & -11.69 & 1.53 & 0.89 & 3,986 \\
20,291 & -10.78 & 1.31 & 0.78 & 4,418 \\
32,470 & -11.24 & 1.30 & 0.78 & 7,233 \\
31,573 & -11.09 & 1.21 & 0.70 & 9,621 \\
34,705 & -11.18 & 1.21 & 0.70 & 10,567 \\
40,950 & -11.51 & 1.33 & 0.80 & 8,330 \\
36,597 & -11.50 & 1.41 & 0.84 & 5,792 \\
41,720 & -11.63 & 1.41 & 0.84 & 6,709 \\
54,096 & -11.96 & 1.46 & 0.86 & 7,391 \\
44,803 & -11.67 & 1.38 & 0.83 & 7,757 \\
37,693 & -11.45 & 1.35 & 0.81 & 7,164 \\
31,629 & -11.12 & 1.23 & 0.72 & 8,898 \\
26,067 & -10.86 & 1.18 & 0.67 & 8,653 \\
25,346 & -10.83 & 1.17 & 0.66 & 8,625 \\
29,435 & -11.06 & 1.24 & 0.72 & 8,103 \\
23,647 & -10.73 & 1.14 & 0.63 & 8,748 \\
20,704 & -10.63 & 1.18 & 0.67 & 6,871 \\
19,388 & -10.58 & 1.19 & 0.68 & 6,218 \\
15,190 & -10.29 & 1.15 & 0.64 & 5,467 \\
21,601 & -10.92 & 1.37 & 0.82 & 3,885 \\
22,278 & -10.99 & 1.40 & 0.84 & 3,636 \\
20,300 & -10.86 & 1.37 & 0.82 & 3,679 \\
19,159 & -10.72 & 1.31 & 0.78 & 4,218 \\
\bottomrule
\end{tabular}
}

    \end{subtable}
\end{table}
\end{landscape}

\end{document}